\pdfoutput=1
\documentclass[journal=acbcct,manuscript=article]{achemso}

\usepackage[version=3]{mhchem} % Formula subscripts using \ce{}
\usepackage{graphicx}
\usepackage[table,xcdraw]{xcolor}
\usepackage{fancyvrb}
\usepackage{xcolor} 
\usepackage{amsmath,bm}
\usepackage{amsfonts}
\usepackage{hyperref}
\usepackage{xr}
\author{Soumendranath Bhakat}
\email{bhakat@wustl.edu/bhakatsoumendranath@gmail.com}
\phone{+1-3054932620}
\affiliation[lunduniversity]
{Division of Biophysical Chemistry, Center for Molecular Protein Science, Department of Chemistry, Lund 
University, P.O. Box 124, SE-22100 Lund, Sweden}
\affiliation[washu]
{Department of Biochemistry and Molecular Biophysics, Washington University, School of Medicine, St. Louis, United States}

\title[An \textsf{achemso} demo]
  {Collective variable discovery in the age of machine learning: reality, hype and everything in between}

\abbreviations{molecular dynamics, metadynamics, aspartic protease, protein dynamics}

\begin{document}
%%%%%%%%%%%%%%%%%%%%%%%%%%%%%%%%%%%%%%%%%%%%%%%%%%%%%%%%%%%%%%%%%%%%%
%% The manuscript does not need to include \maketitle, which is
%% executed automatically.  The document should begin with an
%% abstract, if appropriate.  If one is given and should not be, the
%% contents will be gobbled.
%%%%%%%%%%%%%%%%%%%%%%%%%%%%%%%%%%%%%%%%%%%%%%%%%%%%%%%%%%%%%%%%%%%%%
\begin{abstract}

Understanding kinetics and thermodynamics profile of biomolecules is necessary to understand their functional roles which has a major impact in mechanism driven drug discovery. Molecular dynamics simulation has been routinely used to understand conformational dynamics and molecular recognition in biomolecules. Statistical analysis of high-dimensional spatiotemporal data generated from molecular dynamics simulation requires identification of few low-dimensional variables which can describe essential dynamics of a system without significant loss of informations. In physical chemistry, these low-dimensional variables often called \textit{collective variables}. Collective variables are used to generated reduced representation of free energy surface and calculate transition probabilities between different metastable basins. However the choice of collective variables is not trivial for complex systems. Collective variables ranges from geometric criteria's such as distances, dihedral angles to abstract ones such as weighted linear combinations of multiple geometric variables. Advent of machine learning algorithms led to increasing use of abstract collective variables to represent biomolecular dynamics. In this review, I will highlight several nuances of commonly used collective variables ranging from geometric to abstract ones. Further, I will put forward some cases where machine learning based collective variables were used to describe simple systems which in principle could have been described by geometric ones. Finally, I will put forward my thoughts on artificial general intelligence and how it can be used to discover and predict collective variables from spatiotemporal data generated by molecular dynamics simulations. 


\end{abstract}

\section{Introduction}

Over the past three decades, the major focus of structural biology and biophysics has been to understand conformational dynamics of biological systems across a broad range of timescales with high spatial resolution (ability to visualise molecular motion). Molecular dynamics (MD) simulation acts as a computational microscope which can capture biologically relevant conformational dynamics across varied timescales with high spatial resolution \cite{drormd}. Major advances in computational hardwares and algorithms led to an ever-growing use of MD simulation in varied biological systems. A typical MD simulation generates high-dimensional spatiotemporal data which captures complex molecular motions. Simulation of increasingly larger molecules at ever increasing time scales leads to the `curse of dimensionality' which can be summarised as follows:
``as the size of the biomolecule and the simulation length increases, it also increases the number of explanatory temporal variables (e.g. H-bond distances, radius of gyration, RMSD, dihedral angles etc.) and the problem of structure discovery using temporal variables gets harder. This is analogous to the problem of variable selection during model fitting in machine learning \cite{featuresel}."\\
Capturing low-dimensional representation\footnote{without significant loss of information regarding slow conformational degrees of freedom} from high-dimensional temporal data is an \textit{open} area of research in biomolecular simulation. An excellent example of this is the large scale flap dynamics in plasmepsin-II and BACE-1 \cite{papreview}. In plasmepsin-II, the $sine/cos$ transformation of $\chi_1$ and $\chi_2$ angles of $20$ residues present in the flap region generates $58$ dimensions. Such high number of dimensions makes it incredibly difficult to capture \textit{slow} degrees of freedom which dominates flap \textit{opening}. However, careful analysis of the probability distributions showed that \textit{flipping} of $\chi_1$ and $\chi_2$ angles\footnote{from $58$ to $4$ dimensions} of conserved tyrosine (Tyr) governs the flap dynamics in plasmepsin-II and BACE-1 \cite{bhakatflap}. Such low-dimensional projections which best captures the conformational changes are known as collective variables (CVs) or order parameters\footnote{not to be confused with NMR order parameter which is a measure of conformational entropy}. Temporal evolution along CVs provide thermodynamic and kinetic informations about conformational changes \cite{metadnaturemethods,tiwaryreview,cvml}. It is worth mentioning that the use of \textit{buzz-word} ``automatic identification" to highlight \textit{new} methods for CV discovery from MD simulation gives a false impression i.e selection of good CV is a streamlined and well-defined process. But in practice, naive application of automated methods leads to uninterpretable and poor choice of CVs. Further such methods are not necessarily \textit{new}, but in most cases direct applications or minor modifications of machine learning algorithms developed for classification (binary classifiers and their variants), signal processing (PCA, TICA, autoencoders etc), time-series prediction (neural networks, hidden Markov model etc.) and natural language processing (LSTM, transformers) etc. Recent years saw a sharp rise in applying machine learning algorithms for CV selection on systems which can be represented by simple geometric CVs e.g. dihedral angles, H-bond, RMSDs etc (highlighted in later sections). \\
Due to overuse of the words `AI/machine learning' and flood of papers reporting new algorithms, the current field of CV discovery in biomolecular simulation is hard to navigate for a beginner or people from other disciplines. In this review we will classify the CV discovery process as following: 1) geometric (dihedral angles, H-bonds, RMSD etc) and 2) abstract (PCA, ICA, neural network etc.). One question that is still up in the air: ``Has the field of CV discovery in biomolecular simulation reached a plateau?". I will provide my opinion on the aforementioned question and will put forward some challenges in order to the test the robustness and general applicability of CV discovery algorithms in context of biomolecular simulation. 

\section{Metastability \& collective variables}

\begin{figure}[h]
\hfill\includegraphics[width=10cm,height=10cm,keepaspectratio]{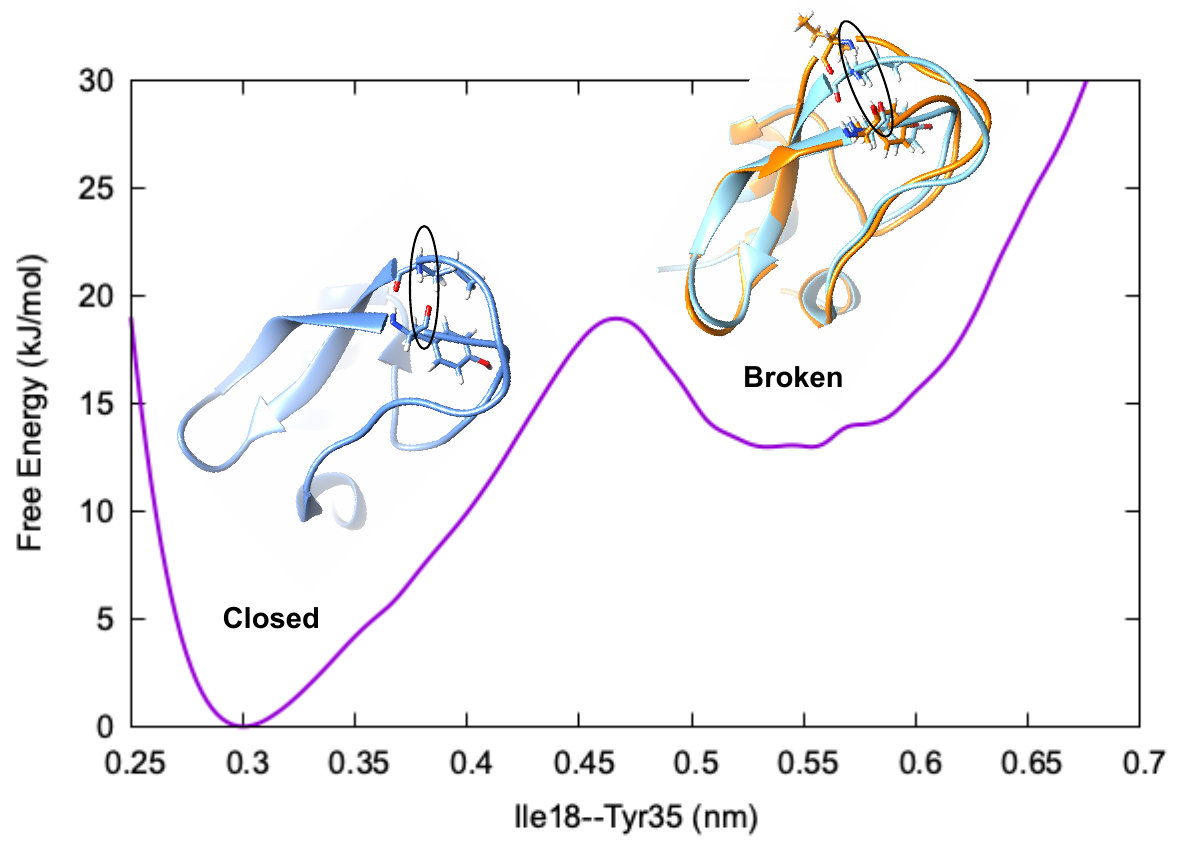}\hspace*{\fill}
\caption{Free energy surface along Ile18--Tyr35 H-bond distance showing two metastable states, closed (blue) and broken (orange) in BPTI. The backbone atoms involved in hydrogen bonding are highlighted in black circle.}
\label{figxhb}
\end{figure}

Before we dive into the definition of CV, we must understand the concept of \textit{metastability}. Let's consider basic pancreatic trypsin inhibitor (BPTI) as an example and focus on a particular H-bond between $Ile18-N-O-Tyr35$ (Fig \ref{figxhb}). In crystal conformation the aforementioned H-bond remains in a \textit{closed} conformation. If one performs a short MD simulation, the stationary probability distribution fluctuates around the closed basin, however, careful analysis of 1 $ms$ MD simulation showed formation of \textit{broken} conformations. A CV (in this case the H-bond distance) should be able to distinguish the metastable states (such as broken and closed conformations) and estimate apparent free energy profile along it as described by following equation:
\begin{equation}\label{eq:1}
F(\mathbf{x}) = - k_B T ln P(\mathbf{x})
\end{equation}

where $\mathbf{x}$ is the CV of choice. 

The \textit{rugged} nature of biomolecular energy landscape is by definition characterised by numerous metastable basins. An \textit{optimal} CV is the one for which two metastable states are separated by high free energy barrier (Figure \ref{figx1}). In most cases a single CV is not enough to capture the complexity of conformational landscape. Selection of CVs is an essential step to perform free energy/kinetics calculations and drive pathway based sampling methods e.g. metadynamics \cite{annualreviewsmetad}, steered MD \cite{steeredmd}, umbrella sampling \cite{umbrella} etc.

\begin{figure}[!htb]
\hfill\includegraphics[width=10cm,height=10cm,keepaspectratio]{Figures/metastability-var.png}\hspace*{\fill}
\caption{Fluctuation of three different variables and their corresponding free energy surfaces highlighting the differences in free energy barrier. One can see that the barrier along \textbf{Var1} is significantly higher compared to other two variables. Heuristically if one has to choose an optimum CV from these three variables, \textbf{Var1} will be the CV of choice.}
\label{figx1}
\end{figure}

For the sake of simplicity CVs can be divided into two categories: 1) geometric and 2) abstract. 

\subsection{Geometric CVs}

Most commonly used geometric CVs that captures conformational dynamics in biomolecules are:

a) Distance: In biomolecular simulation two kinds of distances are most commonly used:\\
i) distance between two atoms \\
ii) distance between center of mass (COM) of two group of atoms.\\

\begin{figure}[h]
\hfill\includegraphics[width=10cm,height=10cm,keepaspectratio]{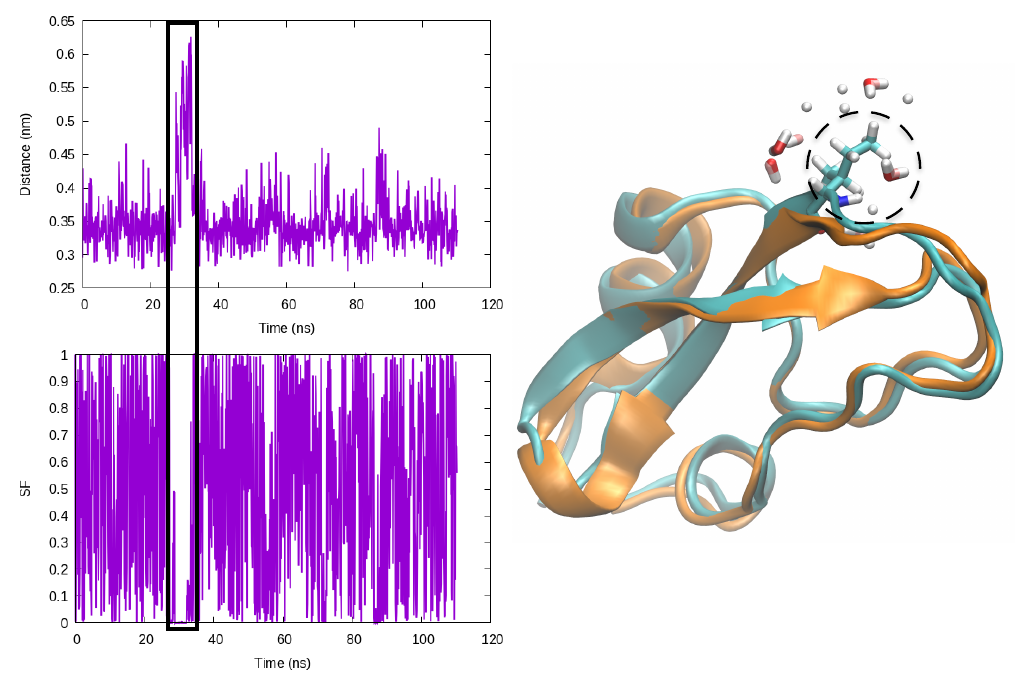}\hspace*{\fill}
\caption{Temporal evolution of H-bond distance of Ile18--Tyr35 and corresponding switching function ($r_{ij}=0.32, r_0=0.06, n=6, m=12$) in BPTI. CV values corresponding broken H-bond conformations is highlighted in black square. Snapshot corresponding broken conformation highlights how breaking of H-bond leads to solvent exposure (highlighting water molecules within 3 $\AA$ radius of Ile18) of Ile18-NH.}
\label{figx2}
\end{figure}

Distance as CVs can be used as an input within adaptive sampling and enhanced sampling methods. For example, distance between COM of ligand and protein can be used as a CV to capture ligand unbinding using well-tempered metadynamics \cite{bhakatflap,Doddameta}. Similarly H-bond distance can be used as CV (Fig \ref{figx2}) to capture transition between closed and broken states. Such CV can be especially useful to study transient solvent exposure in protein which leads to hydrogen-deuterium exchange (paper in preparation).  

b) Switching function: A smoother version of distance CV is \textit{switching} function (SF) which allows a smoother transition between two metastable states along a particular distance (Fig \ref{figx2}). A typical mathematical form\footnote{other functional forms of switching function includes $tanh$, Gaussian, exponential, cubic etc.} of SF can be described as follows:
\begin{equation}\label{eq:2}
s(r)=\frac{ 1 - \left(\frac{ r - r_{ij}}{ r_0 }\right)^{n} }{ 1 - \left(\frac{ r - r_{ij}}{ r_0 }\right)^{m} }
\end{equation}
where $r$  is the instantaneous distance between atoms $i$ and $j$ and $r_{ij}$ is the minimum distance between atoms $i$ and $j$.  For $r<r_{ij}$, $s=1.0$ while $r>r_{ij}$ the function decays smoothly to $0$ (zero). $r_0$ is the value of distance where $s=0.5$. $n$ and $m$ are the hyper-parameters which decides the steepness of the function. \\
A more general CV which combines multiple distances with switching function is known as \textit{contact map}. It calculates the distances between a number of pairs of atoms and convert each distance by a switching function. Such CVs are useful where fluctuation along multiple distances governs conformational dynamics. 

\begin{figure}[h]
\hfill\includegraphics[width=12cm,height=12cm,keepaspectratio]{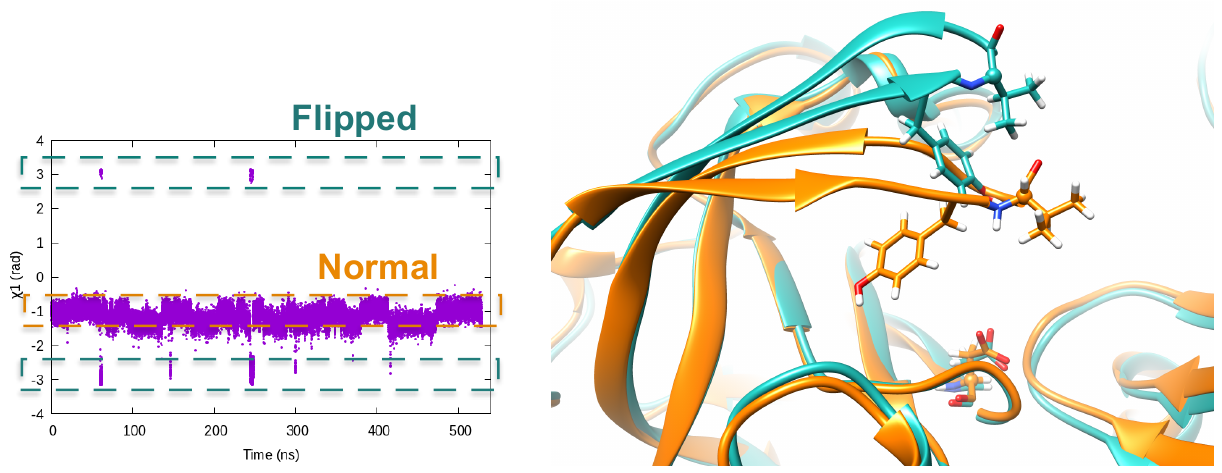}\hspace*{\fill}
\caption{Time-series projection of $\chi1$ angle of Tyr77 in plasmepsin-II during unbiased molecular dynamics simulation (PDB: 1LF4). Tyr77 predominantly remains in the \textit{normal} state (orange) with rare sampling of the \textit{flipped} state (blue). Conformational snapshots corresponding normal and flipped states with relative position of flap ($\beta$ hairpin structure), flap tip residue Val78 and catalytic Asp34 is also highlighted. In plasmepsin-II, rotation of Tyr77 along $\chi1$  and $\chi2$ angles dictates the extent of flap opening which governs substrate entry and catalytic activity.}
\label{figx3}
\end{figure}

c) Dihedral angle: Tracking temporal evolution of dihedral angles (e.g. $\phi, \psi, \omega, \chi_1, \chi_2$) during MD simulation is a well established method to capture conformational dynamics of biomolecules. Most common example includes tracking the conformational sampling in alanine dipeptide by probability distribution along $\phi$ and $\psi$ angles. Recently, \textit{Bhakat and S\"oderhjelm}~\cite{bhakatflap} showed that the conformational dynamics of pepsin-like aspartic proteases can be captured by two dihedral angles\footnote{way to measure similarity between two dihedral CV is to use a function $s = \frac{1}{2} \sum_i \left[ 1 + \cos( \chi_1- \chi_2 ) \right]$ where $\chi_1$ and $\chi_2$ are the dihedral angles and their corresponding instantaneous values.}: $\chi1$ and $\chi2$ angles of conserved Tyr (Fig \ref{figx3} highlights evolution of $\chi1$ during molecular dynamics simulation). However in many cases, conformational dynamics of biomolecules can't be captured by a few dihedral angles. In complex cases, linear combinations of dihedral angles (more on this later) can be used as CVs to capture biomolecular dynamics. 

\begin{figure}[!htb]
\hfill\includegraphics[width=10cm,height=10cm,keepaspectratio]{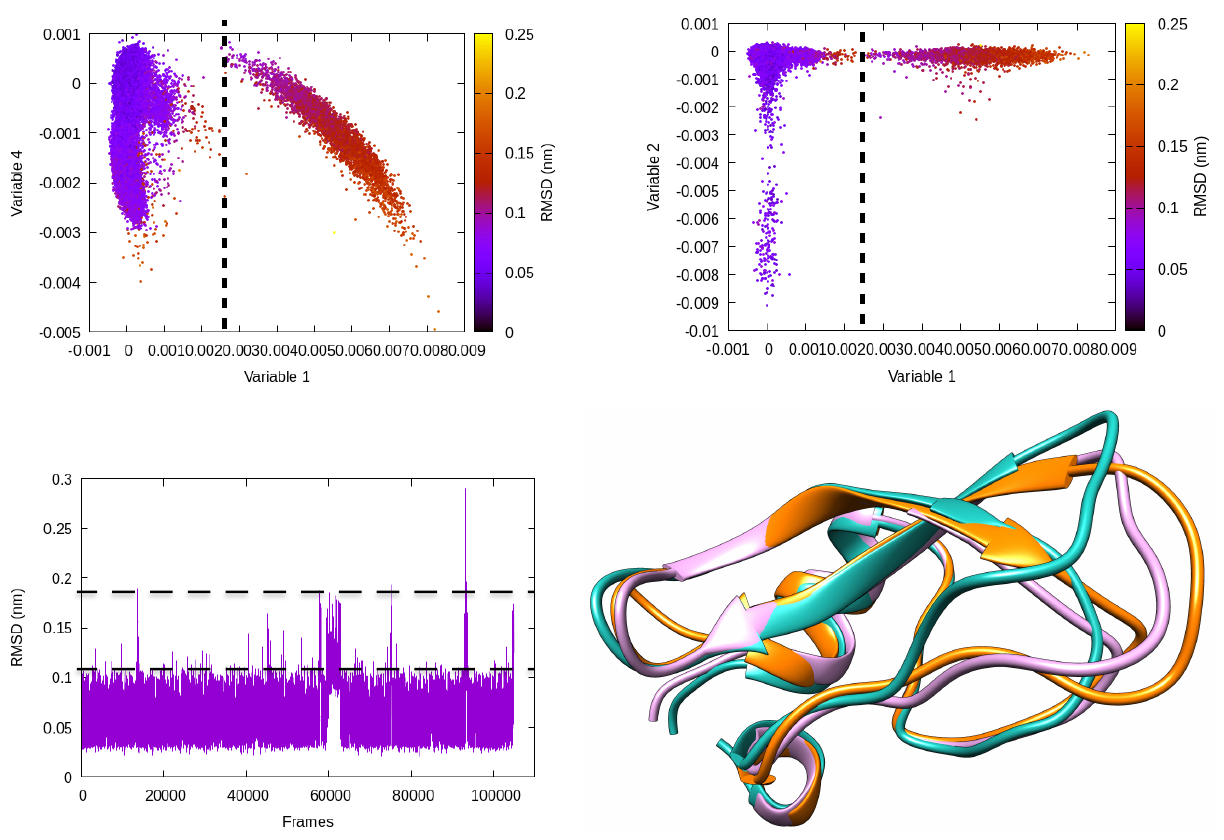}\hspace*{\fill}
\caption{Projection of different variables (more precisely time independent components as described in Figure \ref{figtic1}) which captures transient loop opening BPTI as a function of $C\alpha$ RMSD of residue $6-56$ in BPTI. It can be easily seen that a higher RMSD value separates two metastable states along variable 1. Snapshots corresponding crystal (orange), RMSD$\sim$0.15 nm (pink) and RMSD$\sim$0.25 nm (sea green) highlights differences in loop conformation. RMSD calculation was performed on $M1$ state~\cite{bptim1} of \textit{1 ms} long conventional MD simulation performed by D.E. Shaw research \cite{desres}.}
\label{figx4}
\end{figure}

d) RMSD: RMSD is one of the commonly used CV which measures the similarity between two superimposed atomic co-ordinates. RMSD is the measure of average distance and in bimolecular simulation it measures deviation of atomic co-ordinates from the starting conformation using the following equation:
\begin{equation}\label{eq:3}
RMSD=\sqrt{\frac{1}{n}\sum_{i=1}^{n}d_{i}^{2}}
\end{equation}

where $d_i$ is the distance between atom $i$ and a reference structure. RMSD is usually calculated for $C\alpha$/backbone atoms of the entire protein or for a specific subset. \textit{Stock and coworkers} argued that RMSD is not an optimal choice to capture local conformational (e.g. loop dynamics, domain motions etc.) changes at long-timescales \cite{pca3}. In our unpublished study we have shown that RMSD analysis on a carefully chosen subset can able to capture conformational changes at longer timescales (Figure \ref{figx4}). RMSD based CVs can be used within enhanced sampling and adaptive sampling frameworks to accelerate sampling of the conformational space. 

Besides the aforementioned CVs, several other geometric variables (radius of gyration, $eRMSD$) are frequently used to analyse MD simulations. Softwares e.g. \textit{Plumed} \cite{plumed2}, \textit{gmx} plugins integrated with Gromacs, \textit{CPPTRAJ}~\cite{cpptraj}, \textit{MDAnalysis}~\cite{mdanalysis}, \textit{MDTraj}~\cite{mdtraj} have build-in capabilities to perform analysis of MD trajectories using geometric CVs. 

\subsection{Abstract CVs}

Abstract CVs are usually linear or non-linear transformations of geometric CVs (e.g. dihedral angles, distances, RMSD etc). However the former is often less \textit{intuitive} compared to the latter. In case of linear transformation, the transformed data is a linear combination of original variables. Whereas non-linear transformations are more complex than that. In this section, I will underscore some of abstract CVs (\textbf{linear}: principal component analysis \& independent component analysis; \textbf{non-linear}: \textit{kernel} trick, diffusion map, t-SNE) that have been regularly used to capture low-dimensional spatiotemporal representation from high dimensional dataset generated by MD simulation. I will further discuss the probabilistic interpretation of variational autoencoders which has been applied to capture compressed representation of temporal variables from molecular simulation. Finally, I will highlight the application of binary classifiers in context of classifying metastable states and drive enhanced sampling simulations to capture state transitions. I will further list the softwares/tools/codes that can be used to generate abstract CVs in context of biomolecular simulation. \\

\subsubsection{Principal Component Analysis}
Principal component analysis (PCA) is an unsupervised dimensionality reduction method which transforms a set of variables $r_1, r_2, r_3, ...,r_N$ (where $\mathbf{r} = [r_1, r_2, ...., r_N]^{T}$) to low dimensional representations $y_i$  which captures as much of the variations as possible. It has been widely used by biomolecular simulation community to capture molecular motions with largest amplitude (high variance) cite{pca1,pca2,pca3}. Let $\mathbf{r} \in \mathbb{R}^{D}$ be a vector of geometric CVs e.g. dihedral angles or distances\footnote{assuming $r$ is mean free}. \textit{First} principal component (PC1) of the dataset $r_1, r_2, r_3, ...,r_N$ is the weighted linear combination of the features that captures the largest variance:
\begin{equation}\label{eq:4}
y_1=w_{11}r_1+w_{21}r_2+w_{31}r_3+...+w_{N1}r_N
\end{equation}
where $w_1$ is the vectors of weights or coefficients\footnote{PCA aims to find $w$ which maximises the variance} with elements $w_{11}, w_{21}, ....,w_{N1}$. The elements are normalised i.e. $\sum_{i=1}^{N}{w_{i1}^{2}} = 1$. The process of generating PCs can be summarised as follows:\\
a) generate a \textit{mean free} version of the input data i.e. a vector of geometric CVs.\\
b) compute \textit{eigenvectors} and \textit{eigenvalues} from the covariance matrix 
\begin{equation}\label{eq:5}
\mathbf{C} = \frac{1}{N}\hat{\mathbf{r}}^{T}\hat{\mathbf{r}}
\end{equation}
where $\hat{\mathbf{r}}$ is the mean free version of $\mathbf{r}$.\\
c) sort the eigenvalues in descending order and retain $k$ ($k$ is the new subspace $k < D$) eigenvectors that corresponds to $k$ largest eigenvalues. The eigenvector corresponds to the largest eigenvalue capture the greatest variability in the data.\\
d) construct a projection matrix $\mathbf{w}$ with elements $w_{1j}, w_{2j}, w_{Nj}$ (where $\mathbf{w} = [w_{1j}, w_{2j}, ...., w_{Nj}]^{T}$).\\
e) generate the PCs, $\mathbf{y}$ by transforming the original geometric vectors $\mathbf{r}$ via elements of the projection matrix $\mathbf{w}$.\\

\begin{figure}[h]
\hfill\includegraphics[width=10cm,height=10cm,keepaspectratio]{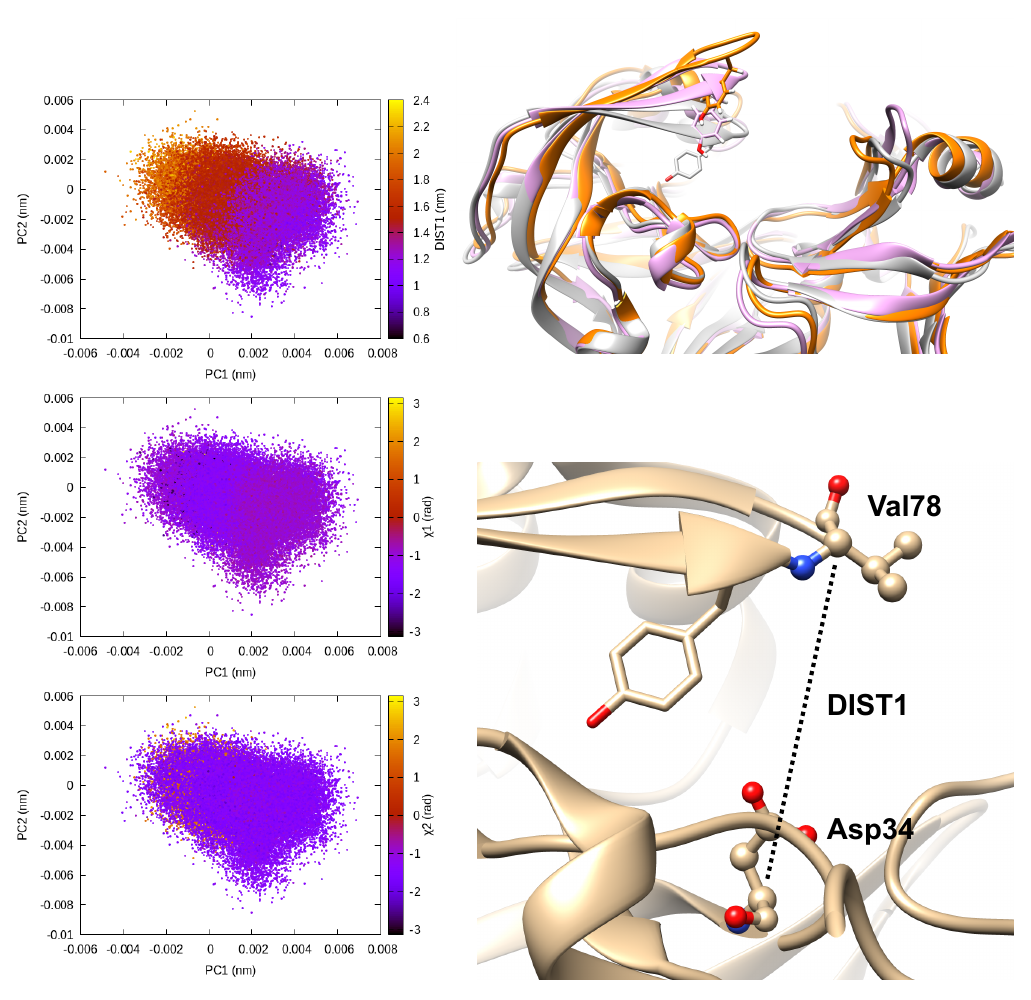}\hspace*{\fill}
\caption{Principal components project along $C\alpha-C\alpha$ distance (DIST1) between Asp34-Val78 and $\chi1$ and $\chi2$ angles of flap tip Tyr77 in plasmepsin-II. One can see that PCs managed to capture the extent of flap opening which is quantified by DIST1 ($DIST1>1.8 nm$ is open conformation). In this case PCA analysis was performed on the the $C\alpha$ atoms of the protein hence it is unable to capture rotational degrees of freedom along $\chi1$ and $\chi2$ angles. Conformational snapshots corresponding crystal (grey, $DIST2=1.2 nm$), semi-open (magenta, $DIST1=1.6 nm$) and open conformation (orange, $DIST1=2.0 nm$) are also highlighted.}
\label{figx5}
\end{figure}

Principal components are \textit{uncorrelated} to each other (as eigenvectors are \textit{orthogonal} to one another) and sorted by their variance ($PC1 > PC2 > PC3 ....$). PCs extracted from MD dataset defines direction in feature space along maximal variance (largest amplitude/fluctuations). PCA has been applied routinely on periodic (dihedral angles\footnote{using $sin$ and $cos$ transformation}) and non-periodic ($CA$ atomic positions, RMSDs, distances) degrees of freedom (Fig \ref{figx5}). The coefficients $w$ corresponding each variables allows incorporating PCs as CVs within enhanced sampling protocols e.g. metadynamics and its variants. PCA reduced dimensions can be also used to construct free energy surface of biomolecules. Despite its success, application of PCA in biomolecular simulation has been a subject of controversy. It has been argued that PCA can't capture the slow conformational degrees of freedom within time scales accessible to MD simulations. However, the author believes that PCA on a carefully chosen subset of atomic co-ordinates \footnote{still an open question and requires further study} can capture slow conformational changes in biomolecules. 

\textbf{Softwares}: \textit{MDAnalysis}, \textit{Plumed} (with capability of using PCs as CVs in metadynamics), \textit{MSMBuilder} \cite{msmbuilder}, \textit{PyEMMA} \cite{pyemma}, \textit{pytraj}, \textit{scikit-learn} (can't be directly used on MD trajectories but can be interfaced with MD post-processing tools e.g. \textit{MDTraj}).

\subsubsection{Independent Component Analysis}

Independent component analysis is a dimensionality reduction method which transforms a set of vectors (e.g distances, RMSDs, dihedral angles etc.) into maximally independent linear combinations (independent components). Imagine $\mathbf{y}(t) = [y_1(t), ......., y_N(t)]^{T}$ is a linear mixture of high-dimensional data $\mathbf{r}(t) = [r_1(t), ......, r_N(t)]^{T}$\footnote{data is mean free}  such that $\mathbf{y}(t)=\mathbf{A}\mathbf{r}(t)$ where $\mathbf{A}$ is the square mixing matrix and the components of $\mathbf{y}(t)$ are mutually independent. Two components can said to be independent if their joint distribution is equal to the product of their marginals\footnote{a corresponding measure of independence is Kullback-Leibler divergence}:
\begin{equation}\label{eq:ic1}
p(y_1(t),y_2(t)) = p(y_1(t))p(y_2(t))
\end{equation}

where $p(y_1(t),y_2(t))$ are the joint probability distributions and $p(y_1(t))p(y_2(t))$ is the marginal along two components $y_1(t)$ and $y_2(t)$.  However the aforementioned definition of independence has a drawback: lets consider a signal without time auto-correlation and a second signal which is equal to the first signal but shifted in time (often known as signal with time-delay). If one applies equation \ref{eq:ic1} the two signals will appear to be mutually independent. The time-delayed signal can be called statistically independent if all the time-delayed correlations are zero (second-order ICA) \cite{schustertica}. In biomolecular simulation second-order ICA is known as time-lagged independent component analysis (TICA) \cite{noetica,pandetica}. The first step of second-order ICA/TICA is to introduce a time-lagged (or time-delayed) correlation matrices of the input variables $\mathbf{r}(t)$:
\begin{equation}\label{eq:ic2}
\mathbf{C}^{r}(\tau)=< \mathbf{r}(t)\mathbf{r}(t+\tau)^{T}>
\end{equation}
where $\tau$ is the lag-time or time delay between two signals. The entries of $\mathbf{C}^{r}(\tau)$\footnote{should be diagonal for all $\tau$} are denoted as $\mathbf{C}_{ij}^{r}(\tau)$. The common practice is to express $\mathbf{C}^{r}(\tau)$ as a symmetrized version of correlation matrices:
\begin{equation}\label{eq:ic3}
\mathbf{C}^{r}(\tau)=\frac{1}{2}\left [ < \mathbf{r}(t)\mathbf{r}(t+\tau)^{T}> + < \mathbf{r}(t+\tau)\mathbf{r}(t)^{T}> \right ]
\end{equation}
Symmetrization is a mathematical trick which allows applying the algorithm to the reversible dynamics (as $< \mathbf{r}(t+\tau)\mathbf{r}(t)^{T}> = <\mathbf{r}(t)\mathbf{r}(t-\tau)^{T}>$). It also makes sure that the eigenvalues are \textit{real} and two eigenvectors that comes from distinct eigenvalues are \textit{orthogonal}. Finally the TICA problem can be formulated a generalised eigenvalue problem\footnote{can be solved by second-order blind source separation algorithms e.g. AMUSE}:
\begin{equation}\label{eq:ic4}
\mathbf{C}^{r}(\tau)\:\mathbf{A}=\mathbf{C}^{r}(0)\:\mathbf{A}\:\boldsymbol{\Lambda}
\end{equation}
where $\mathbf{A} = (a_1,....a_N)$ is the orthogonal matrix of generalised eigenvectors and $\boldsymbol{\Lambda}=diag(\lambda_1,....\lambda_N)$ is the diagonal matrix of eigenvalues\footnote{the eigenvalues $\lambda_1,....\lambda_N$ are associated with eigenvectors of $a_1,....a_N$.}. $\mathbf{A}$ contains the independent components (ICs) the original data $\mathbf{r}(t)$ can be projected on the TICA space as: $\mathbf{y}(t)=\mathbf{A}\mathbf{r}(t)$. The scalar components of $\lambda_i$ captures the magnitude of the auto-covariance where smaller $\lambda_i$ captures largest auto-covariance:
\begin{equation}\label{eq:ic5}
\begin{aligned}
\lambda_1 < \lambda_2 < .....<\lambda_N \scriptstyle{\text{; } \lambda_1 \text{ captures largest auto-covariance and so on}}
\end{aligned}
\end{equation}

The eigenvalues, $\lambda_i$ are associated with the relaxation timescales of a biomolecular process by:
\begin{equation}\label{eq:ic6}
t_i = -\frac{\tau}{\ln |\lambda_i|}
\end{equation}

\begin{figure}[h]
\hfill\includegraphics[width=10cm,height=10cm,keepaspectratio]{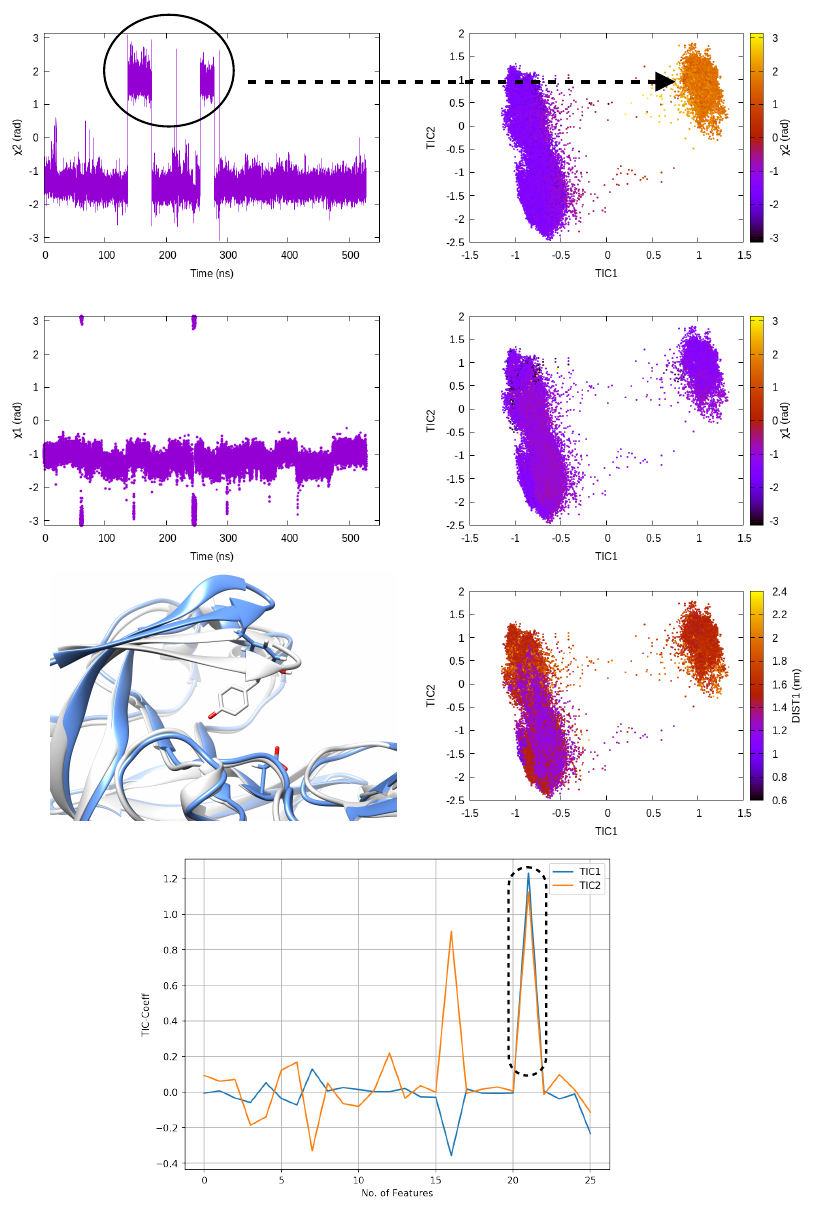}\hspace*{\fill}
\caption\protect{Evolution of $\chi1$ and $\chi2$ angles of Tyr77 of plasmepsin-II during MD simulation. TICA analysis\footnote{with lag time 1000} was performed on $\chi1$ and $\chi2$ angles ($sin/cos$ modification) of residue 74-84. Projection of TIC1 and TIC2 along $\chi1$ and $\chi2$ angles of Tyr77 shows that TIC1 managed to capture the slowest degree of motion which is the $\chi2$ rotation of Ty77. Further, projection along TICs failed to separate fluctuation along DIST1 as it is defined by distance between backbone $C\alpha$ atoms. Fluctuation of TIC coefficients (eigenvalues) also highlights feature no $21$ ($sin\chi2$ of Tyr77) as the dominant feature among $25$ features (Table 1 in Supplementary Informations).  Snapshots corresponding crystal (grey) and flipped ($\chi2\sim$ 2 rad) are also highlighted.}
\label{figx7}
\end{figure}

\begin{figure}[h]
\hfill\includegraphics[width=10cm,height=10cm,keepaspectratio]{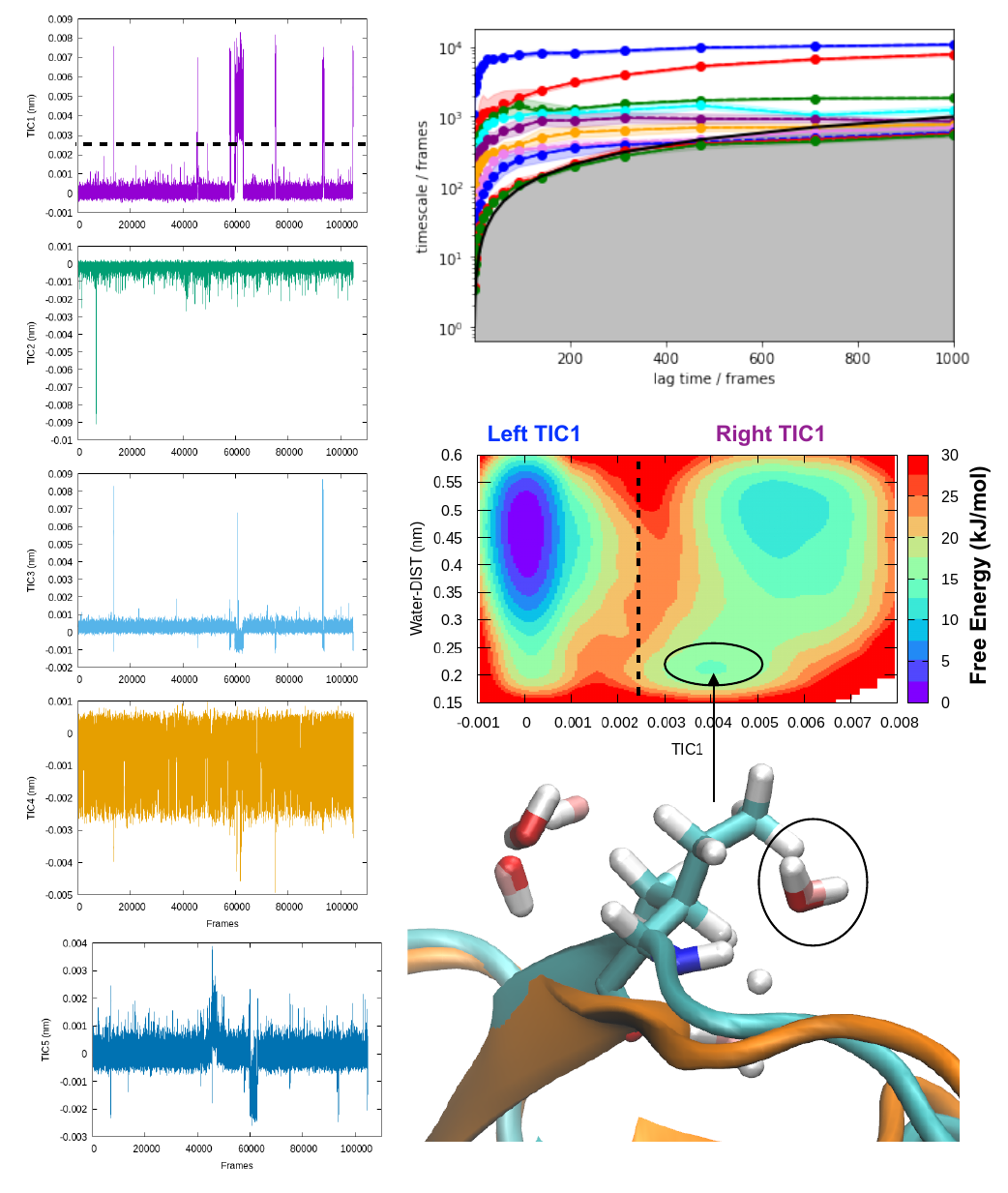}\hspace*{\fill}
\caption{Time-series projection of first 5 TICs and corresponding implied timescales showing how first few TICs captured slow conformational degrees of freedom (transient loop opening as depicted in Figure \ref{figx4}) in BPTI. TICA analysis was performed on $C\alpha$ atomic positions of residues $6-56$ using a lag time 500 (all calculations were performed on M1 state as described by ref \cite{bptim1}). Transient loop opening in BPTI leads to breaking of H-bod interaction between $Ile18-NH$ and $Tyr35-O$ which makes the amide of $Ile18$ solvent exposed and enables hydrogen-deuterium exchange \cite{hdxbpti}. Projection of water distance from $Ile18-NH$ shows how TIC1 captures a metastable solvent exposed basin where a water molecule comes within $0.25$ nm of amide hydrogen. Snapshot corresponding solvent exposed conformation highlights the relative position of backbone amide and closest (with $0.3$ nm) waters.}
\label{figtic1}
\end{figure}

TICA has been routinely used on temporal data from molecular dynamics simulation to identify slow conformational degrees of freedom such as flipping of side-chain dihedral angle (Figure \ref{figx7}), transient exposure of protein interior which leads of solvent exposure (Figure \ref{figtic1}) and drive enhanced sampling simulations \cite{ticametad}. Recently Schultze and Grubmuller compared projection of TICs between high-dimensional data from protein dynamics and random walk \cite{schultze}. However the authors disregarded any discussion surrounding the choice of input features and its effect on TICA. Figure \ref{figx7} highlights the importance of input features in describing conformational dynamics along TIC space. Lack of \textit{separation aka population gap} along TIC space in ref \cite{schultze} (see Figure 7, left panel) is a sign that either TICA analysis was performed on a poorly chosen feature space or the sampling was not sufficient enough to capture slow conformational dynamics in proteins.

\begin{figure}[h]
\hfill\includegraphics[width=10cm,height=10cm,keepaspectratio]{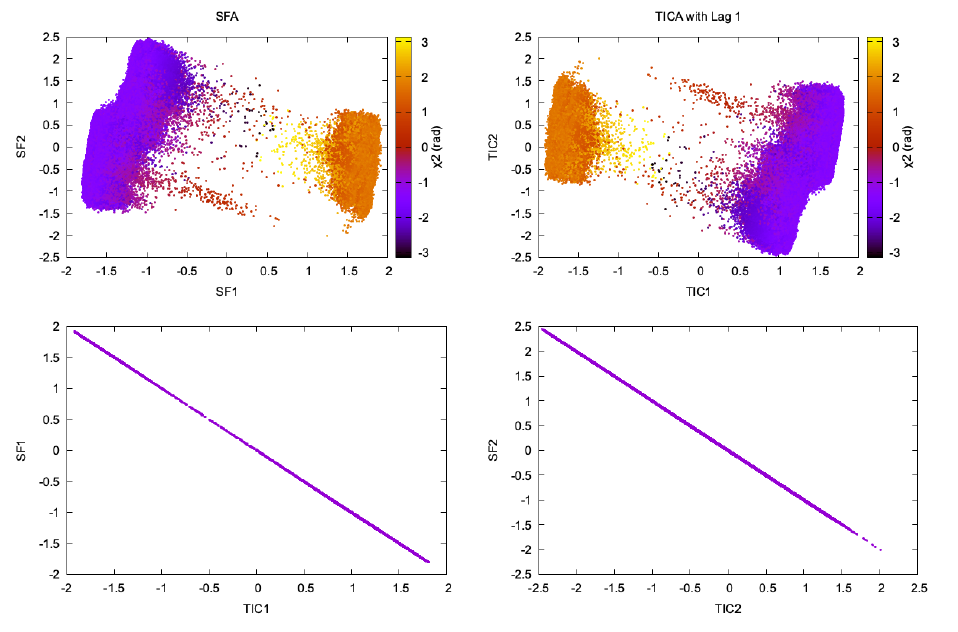}\hspace*{\fill}
\caption{Projection of slow features and TICs with lag time $1$ along $\chi2$ angle of Tyr77 shows similarity of TICA and SFA in separating slow conformational degree of freedom. The squared Pearson correlation coefficient, $R^2$ between SFA and TICA with lag time $1$ is $1.00$. This example is a demonstration of mathematical concepts described in ref \cite{sfavsica} in context of biomolecular simulation. It further raises possibilities of combining the two algorithms and use it to develop Markov state model.}
\label{figtic2}
\end{figure}

Wiskott and co-workers \cite{sfavsica} have shown how the objective function of TICA with lag time $1$ is formally equivalent to \textit{slow feature analysis} (SFA)\footnote{captures slowly varying features from high-dimensional input} (Figure \ref{figtic2}). SFA and TICA are based on two different principles: \textit{slowness} and \textit{statistical independence}. However, the similarity between SFA and TICA (second order ICA with lag-time $1$) opens up possibilities to combine these two algorithms~\cite{sfaandica} for capturing \textit{slowly} varying \textit{statistically independent} components from high-dimensional temporal data generated by MD simulation. 

\textbf{Softwares}: \textit{MSMBuilder}~\cite{msmbuilder}, \textit{PyEMMA}~\cite{pyemma} and \textit{deeptime}~\cite{deeptime} comes with built in TICA functionality. A scikit-learn style implementation of SFA can be found here: \url{https://github.com/wiskott-lab/sklearn-sfa} .

\subsubsection{Kernel trick}

MD simulation generates complex and non-linear representation of biomolecular dynamics. \textit{Kernel} trick is a mathematical transformation which maps the original data into a higher dimensional feature space which is then used to find linear projections using PCA (kernel PCA) or tICA (kernel TICA \cite{kerneltica}). For data points $r_i, r_j$ in the input space $\mathbb{N}$, kernel function $k(r_i,r_j)$ generates modified inner products which maps $\mathbb{N}\rightarrow \mathbb{Z}$
\begin{equation}\label{eq:ic7}
k(r_i,r_j)=< \phi (r_i), \phi(r_j)> _\mathbb{Z}
\end{equation}
where $\phi$ is the mapping function and $<.,.>_\mathbb{Z}$ must be proper inner product. One doesn't need to explicitly compute $\phi$ as the kernel matrix $k(r_i,r_j)$ (modified inner product) can be easily computed by a variety of functions: \\
a) \textit{Polynomial} kernel: $k(r_i,r_j)=(\gamma . r_i^{T}r_j+c)^{d}, \gamma>0$ \\
b) \textit{Sigmoid} kernel: $k(r_i,r_j)=tanh(\gamma . r_i^{T}r_j+r)$ \\
c) \textit{Gaussian} kernel: $k(r_i,r_j)=exp(-\gamma . \left \|  r_i-r_j\right \|^{2}), \gamma>0$

where $r$, $d$ and $\gamma$ are kernel hyper-parameters. A drawback of this approach is that when we map the data into higher dimensions, we may overfit the model. Hence the choice of right kernel functions are of utmost importance. Schwartz and colleagues used kernel PCA (using polynomial kernel function) to identify CVs in lactate dehydrogenase. Further, kernel TICA has been used to identify CVs which captures folded to unfolded transition in small folded proteins \cite{landmarkkernel}. 

\textbf{Softwares}: \textit{scikit-learn} and \textit{MODE-TASK}~\cite{modetask} have in-built kernel PCA functionality. \textit{MSMBuilder} has a built in kernel TICA algorithm which can be applied on high-dimensional features from MD dataset for discovering low-dimensional representations \textit{aka} CVs. 

\subsubsection{Diffusion Map}

Diffusion map is a non-linear dimensionality reduction technique. It combines the concept of random walk Markov chain and diffusion process by projecting the input data in a low-dimensional space where the distance (e.g Euclidean) between data points resembles the diffusion distance in the original high-dimensional space. When applied to MD dataset, diffusion map generated vector space (CVs) are constructed in such a way that conformations which are kinetically closed are placed together whereas kinetically distant (separated by high free-energy barrier) conformations are placed far apart. \\
In a random walk Markov model the connectivity between data points $r_i$ and $r_j$ is defined as the probability of jumping from state $r_i$ to $r_j$ in one step of random walk. The connectivity can be also expressed as a normalised kernel function $k(r_i,r_j)$\footnote{similar functional form of Gaussian kernel. One can choose other measures of distances as kernel functions e.g. Mahalonibis distance, Jensen-Shannon divergence etc.} which measures similarity between data points:

\begin{equation}\label{eq:dm1}
\begin{aligned}
k(r_i,r_j)= exp(-\gamma . \left \|  r_i-r_j\right \|^{2}), \gamma>0 \\
= exp(-\frac{\left \|  r_i-r_j\right \|^{2}}{\alpha})
\end{aligned}
\end{equation}

where the value of $\alpha$ decides the size of neighbourhood. It can also be seen as a hyper-parameter which chooses the conformations that are kinetically closed. In practice conformations closer than the value of $\alpha$ are relevant for $k_{ij}=k(r_i,r_j)$ as the contribution from distant conformations decays exponentially. The diffusion kernel satisfies the following two properties:\\
a) $k$ is symmetric: $k(r_i,r_j)=k(r_j,r_i)$. This allows spectral analysis of the distance matrix $k_{ij}$\\
b) $k(r_i,r_j)\geq 0$. \\

The transition probability $p(r_i,r_j)$ can be expressed as:
\begin{equation}\label{eq:dm2}
\begin{aligned}
p(r_i,r_j)=\frac{1}{D}k(r_i,r_j)
\end{aligned}
\end{equation}

where $D$ is the normalisation factor. The normalised transition matrix $\mathbf{P}$ with $P_{ij}=p(r_i,r_j)$ (where $\sum_{j} P_{ij} = 1$) allows spectral (in other words \textit{eigen}) decomposition:
\begin{equation}\label{eq:dm3}
\begin{aligned}
\mathbf{P}\mathbf{A}=\boldsymbol{\Lambda}\mathbf{A}
\end{aligned}
\end{equation}

where $a_1,....,a_N$ ($\mathbf{A} = (a_1,....a_N)$) are real valued eigenvectors corresponding to eigenvalues $\lambda_1,....,\lambda_N$ ($\boldsymbol{\Lambda}=diag(\lambda_1,....\lambda_N)$ ) where $1=\lambda_1 > \lambda_2  ... \geq  \lambda_N$.  Similarly to TICA, the eigenvalues generated from normalised transition matrix $\mathbf{P}$ shows \textit{spectral gap} which can be expressed as the difference between two largest eigenvalues $\lambda_1 - \lambda_2$ or more generally $1-max\left \{ |\lambda_2|,|\lambda_N| \right \}$. Diffusion map at time $t$ can be approximated as mapping between the original space and the latent space of first $k$ eigenvectors:
\begin{equation}\label{eq:dm4}
\begin{aligned}
\mathbf{A}(\mathbf{r})=(\lambda_{1}^{t}a_1(r),\lambda_{2}^{t}a_2(r),...,\lambda_{k}^{t}a_1(r))
\end{aligned}
\end{equation}

The diffusion distance\footnote{analogous to the functional form of kinetic distance as proposed by Noe and Clementi ref \cite{Noeclementi}} at time $t$ can be expressed as a function of eigenvectors. Kevrekidis and coworkers \cite{dmapneuralips} has shown that the diffusion distance can be approximated as Euclidean distance on the diffusion map space:
\begin{equation}\label{eq:dm5}
\begin{aligned}
D_{t}^{2}(r_i,r_j)=\sum_{l}\lambda_{l}^{2t}\left ( a_l(r_i)-a_l(r_j) \right )^{2}=\left \| \mathbf{A}_{t}(r_i)-\mathbf{A}_{t}(r_j) \right \|^{2}
\end{aligned}
\end{equation}
Equation \ref{eq:dm4} provides justification of using Euclidean distance in the diffusion map space for clustering. Diffusion distance can also be interpreted as a measure of how kinetically close are the two conformations. The distance is small if there are multiple high probability transition pathways between two conformations. Diffusion map has been applied in biomolecular simulation to capture \textit{slow} transitions and guide enhanced sampling simulations \cite{diffusionmap1,diffusionmap2,diffusionmap3}. Recently the functional form of equation \ref{eq:dm4} in combination with maximum entropy based CV selection method SGOOP\footnote{in principal one can use the same strategy to combine diffusion distance with TICA/SFA} was used to capture kinetically relevant low dimensional projection in small peptides \cite{sgoopd}. 

\textbf{Softwares}: MDAnalysis has an integrated diffusion map module which can be applied on selected features from MD simulation.

\subsubsection{t-distributed stochastic neighbour embedding}

t-distributed stochastic neighbour embedding (t-SNE) is a non-linear dimensionality reduction method which performs embedding of high-dimensional data to low-dimensional space such that neighbouring data-points are assigned highest probability while distant points are assigned lower probability \cite{JMLRtsne,tsnedistill}. For a high-dimensional space $r_1,.....,r_N$ the distance between $r_j$ and $r_i$ can be expressed as:
\begin{equation}\label{eq:tsne1}
\begin{aligned}
p_{j|i}=\frac{exp(- \left \|  r_i-r_j\right \|^{2}/2\sigma_{i}^{2})}{\sum_{k\neq i}exp(- \left \|  r_i-r_k\right \|^{2}/2\sigma_{i}^{2}}
\end{aligned}
\end{equation}
where $p_{j|i}$ is the conditional probability which captures the similarity between data points $r_j$ with $r_i$ such that $p_{i|i}=0$ and $\sum_{j}p_{j|i}=1$. The functional form of equation \ref{eq:tsne1} is similar to Gaussian kernel in equation \ref{eq:dm1}. The conditional probability can be transformed into a joint probability distribution of symmetrized matrix $P_{ij}$:
\begin{equation}\label{eq:tsne2}
\begin{aligned}
P_{ij}=\frac{p_{j|i}+p_{i|j}}{2N}
\end{aligned}
\end{equation}
where $P_{ij}=P_{ji}$, $P_{ii}=0$ and $\sum_{i,j}P_{j|i}=1$.

\begin{figure}[h]
\hfill\includegraphics[width=10cm,height=10cm,keepaspectratio]{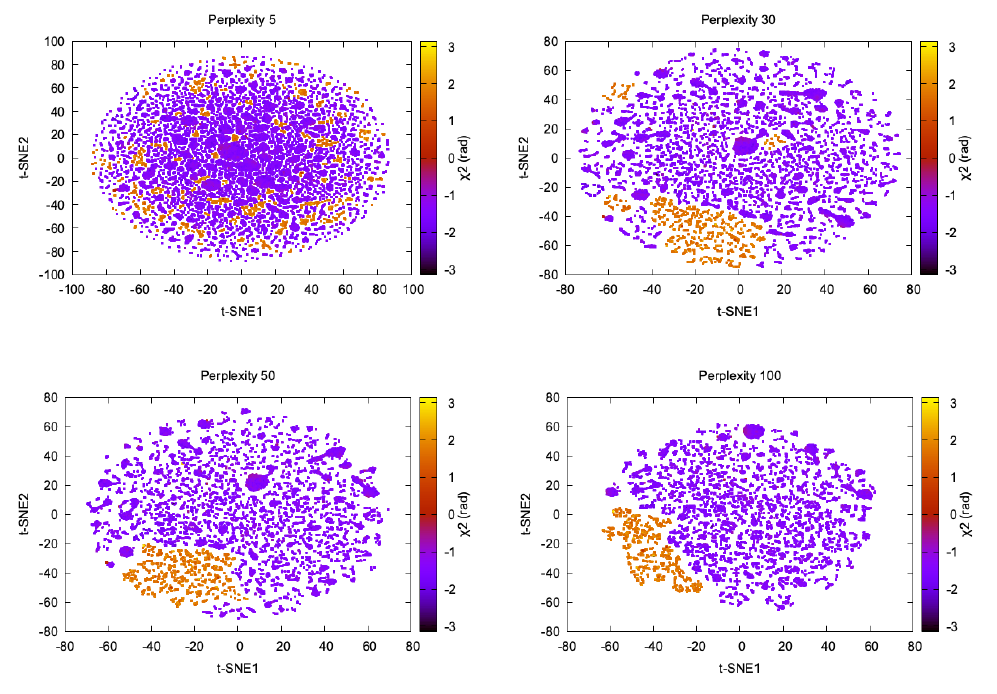}\hspace*{\fill}
\caption{Projection of first two t-SNE components along $\chi2$ angle of Tyr77 in plasmepsin-II shows how the embedding changes with perplexity. Further t-SNE performed on the high dimensional torsional features (Table 1 in supplementary informations) didn't manage to separate flipping along $\chi2$ which is a slow degree of freedom. It further shows that clustering on top of t-SNE reduced dimensions will generate artificial clusters which in reality belong to same metastable state.}
\label{figtsne1}
\end{figure}

The Gaussian bandwidth $\sigma_{i}$ has been set in such a way that the perplexity of the conditional distribution equals to the perplexity provided by the user\footnote{ref \cite{JMLRtsne} highlighted that there is a monotonically increasing relationship between perplexity and bandwidth}. The perplexity is defined as:
\begin{equation}\label{eq:tsne3}
\begin{aligned}
Perp(p_i)=2^{H(p_i)}
\end{aligned}
\end{equation}
where $p_i$ is the conditional probability distribution over all the data-points given $r_i$ and $H(p_i)$ is the Shannon entropy $H(p_i)=-\sum_{j}p_{j|i}log_{2}p_{j|i}$. Perplexity can be thought as a measure of nearest neighbours and the choice of perplexity heavily determines the outcome of t-SNE (Figure \ref{figtsne1}). \\
t-SNE aims to learn a low dimensional manifold $r'_1,...,r'_N$ in such a way that the new conditional probability $p'_{j|i}$ reflects similarity with $p_{j|i}$. $p'_{j|i}$ can be expressed as:
 \begin{equation}\label{eq:tsne4}
\begin{aligned}
p'_{j|i}=\frac{(1+ \left \|  r'_i-r'_j\right \|^{2})^{-1}}{\sum_{l\neq k}(1+ \left \|  r'_k-r'_l\right \|^{2})^{-1}},  p'_{i|i}=0
\end{aligned}
\end{equation}
The distance based metrics $(1+ \left \|  r'_i-r'_j\right \|^{2})^{-1}$  is a heavily tailed distribution. In t-SNE, Student t-distribution has been used to measure the similarity between low-dimensional data-points so that the points that are far apart have $p'_{j|i}$ which are invariant of perturbation. The algorithm aims to minimise the \textit{Kullback-Leibler} divergence\footnote{other measures of similarity e.g. Jensen-Shannon divergence, Bhattacharya distance can be explored to improve the embedding result} (using gradient descent) by comparing joint probability distributions $P'$ (low-dimensional) and $P$ (high-dimensional):
 \begin{equation}\label{eq:tsne5}
\begin{aligned}
KL(P || P')=\sum_{i\neq j}p_{ij}log\frac{p_{ij}}{q_{ij}}
\end{aligned}
\end{equation}
The gradient descent algorithm is usually effective for small datasets but performs poorly in case of larger datasets. Recently a time-lagged version of t-SNE \cite{timelagtsne} was proposed using inspiration from time-lagged ICA (TICA). However, the stochastic nature (mainly due to \textit{perplexity}) of t-SNE (as shown in Figure \ref{figtsne1} and discussed in ref \cite{liortsne}) prevents its use as a meaningful CV for reconstructing trustworthy free energy surface and calculating kinetics from MD simulation\footnote{recently proposed feed forward neural network based tSNE otherwise known as parametric t-SNE can also be applied on MD dataset}. Clustering on the top of t-SNE embedding can also produce artificial clusters which might trick the user of thinking that it discovers new metastable states but in reality they belong to the same free energy basin. 

\textbf{Softwares}: Time-lagged version of t-SNE can be accessed here: \url{https://github.com/spiwokv/tltsne}. Popular machine learning package \textit{scikit-learn} also has a t-SNE module: \url{https://scikit-learn.org/stable/modules/generated/sklearn.manifold.TSNE.html}. MODE-TASK \url{https://github.com/RUBi-ZA/MODE-TASK} a python toolkit for analysing MD simulation has an integrated t-SNE functionality. Tensorboard embedding projector \url{https://projector.tensorflow.org} has a graphical user interface to perform t-SNE. 

\begin{figure}[h]
\hfill\includegraphics[width=10cm,height=10cm,keepaspectratio]{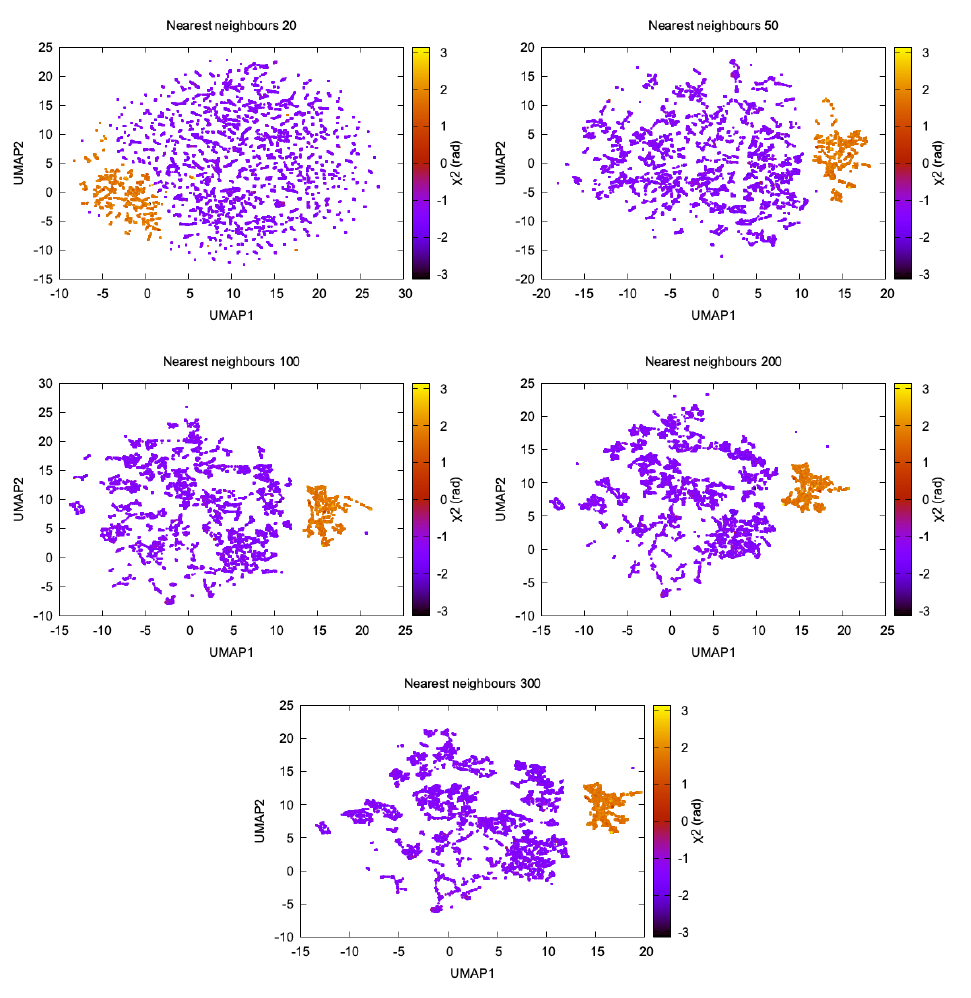}\hspace*{\fill}
\caption{UMAP components projected along $\chi2$ of Tyr77 highlights how the shape of UMAP vary as a function of \textit{nearest neighbour} hyper-parameter. UMAP converges with nearest neighbour $200$. UMAP does a better job compared to t-SNE in separating conformational space along $\chi2$. I believe clustering on top of UMAP reduced dimensions might be useful in understanding conformational heterogeneity within a broad metastable basin.}
\label{figumap}
\end{figure}

In recent years, several other dimensionality reduction methods such as spectral gap optimization of order parameters (SGOOP) \cite{sgoop}, isomap~\cite{mdsisomap} (\url{https://scikit-learn.org/stable/modules/generated/sklearn.manifold.Isomap.html}), dynamic mode decomposition (DMD)~\cite{dmd}(\url{https://mathlab.github.io/PyDMD/}, see ref \cite{datadriven} for similarity between DMD and TICA and their variants), multi-dimensional scaling (MDS)~\cite{mdsisomap}, UMAP~\cite{umap,umapmd} (\url{https://umap-learn.readthedocs.io/en/latest/} , Figure \ref{figumap}), iVIS~\cite{ivis} (\url{https://bering-ivis.readthedocs.io/en/latest/}) which are similar to some of the aforementioned algorithms, were applied on temporal data from MD simulation. Further, \textit{sparse regression} based dimensionality reduction method \textit{sparse identification of nonlinear dynamical systems} (SINDy) \cite{sindy} can possibility be applied on temporal data from molecular dynamics simulation to discover linear combinations of features which best captures conformational dynamics in macromolecules. Recent review by \textit{Glielmo et al} \cite{noe-chemicalreviews} summarises mathematical concepts, strengths and limitations as well as applicability of few such methods in analysing high-dimensional data from MD simulations.  

\subsubsection{Learning co-ordinates for dynamics using variational autoencoder}

Variational autoencoder (VAE) is a dimensionality reduction method which takes high dimensional data (e.g multi-variate geometric or abstract CVs) as inputs and learns latent (compressed) representations that captures minimal essential information necessary to describe the dynamics of the system. VAE is a type of autoencoder whose principals are deeply rooted in \textit{variational} statistics \cite{vaewelling,autoencoding}. In this article we will use the language of \textit{probability theory} to describe VAE and try to establish a connection between free energy and the \textit{loss function} of VAE.  \\
VAE takes high dimensional vector $\mathbf{r} = [r_1, r_2, ...., r_N]^{T}$ as inputs. Each components of $\mathbf{r}$ are probability distributions along geometric or abstract CVs. The encoder part of the VAE encodes the high dimensional data into the latent variables $\mathbf{z}$. The joint probability distribution of input and latent variables, $p(r,z)$ can be expressed as:\\
\begin{equation}\label{eq:6}
p(r,z)=p\left( {r|z} \right)p\left( z \right)
\end{equation}
The generative process can be expressed as:\\
a) sampling latent variables, $z_i$ from the prior distribution $p\left( z \right)$ and \\
b) sampling of data point $r_i$ from the likelihood $p\left( {r|z} \right)$ which is conditional on the latent variables $z$.\\

The goal of VAE is to infer good approximation of the latent variables ($z$) given input data ($r$) which is equivalent to calculate the posterior distribution $p\left( {z|r} \right)$ as following:\\
\begin{equation}\label{eq:7}
p\left( {z|r} \right) = \frac{{p\left( {r|z} \right)p\left( z \right)}}{{p\left( r \right)}} = \frac{p(r,z)}{p\left( r \right)}
\end{equation}

where
\begin{equation}\label{eq:8}
p\left( r \right)= \int {p\left( {r|z} \right)p\left( z \right)dz}
\end{equation}

Estimation of $p\left( r \right)$ is a computationally expensive process as it requires integrating over all the possible values of $z$. VAE approximates the posterior distribution $p\left( {z|r} \right)$ by a new distribution $q\left( {z|r} \right)$ (tractable distribution). If $q_\lambda\left( {z|r} \right)$ is similar to $p\left( {z|r} \right)$  then we can use it to approximate $p\left( r \right)$. $\lambda$ is the hyper-parameter which indicates type of distribution. If $q\left( {z|r} \right)$ a Gaussian distribution then $\lambda_{r_i}=(\mu_{r_i}, \sigma ^{2}_{r_i} )$ where $\mu$ is the mean and $\sigma ^{2}$ is the variance of latent variables of each input component. \\
Now we want to measure how well $q_\lambda\left( {z|r} \right)$ approximates $p\left( {z|r} \right)$. VAE uses Kullback-Leibler (KL) divergence to measure information loss between two probability densities as described below:
\begin{equation}\label{eq:9}
\begin{aligned}
KL(q_{\lambda}( {z|r} ) || p ({z|r})) = \mathbb{E}_{q} \left [ log\frac{q_{\lambda}( {z|r} )}{p ({z|r})} \right ] \\
= \mathbb{E}_{q} \left [ log q_{\lambda}( {z|r} ) - log p ({z|r}) \right ] \\
= \mathbb{E}_{q} \left [ log q_{\lambda}( {z|r} ) \right ] - \mathbb{E}_{q} \left [ log p(r, z) \right ] + log p(r) \scriptstyle{\text{; taking } log \text{ of eq. \ref{eq:7}}}\\
= - \left ( \mathbb{E}_{q} \left [ log p(r, z) \right ] -  \mathbb{E}_{q} \left [ log q_{\lambda}( {z|r} ) \right ]  \right ) + log p(r) \\
= - ELBO(\lambda) + log p(r)
\end{aligned}
\end{equation}

We can reformulate equation \ref{eq:9} as follows:
\begin{equation}\label{eq:10}
log p(r) = ELBO(\lambda) + KL(q_{\lambda}( {z|r} ) || p ({z|r}))
\end{equation}

From equation \ref{eq:10} we can see that minimising KL divergence (KL divergence is always greater or equal to zero) is equivalent to maximising Evidence Lower Bound (ELBO). It allows us to bypass the hard task of minimising KL divergence between the approximate $q( {z|r} ) $ and true posterior $p ({z|r})$ , instead we maximise ELBO. ELBO term can be further expressed as follows:
\begin{equation}\label{eq:11}
\begin{aligned}
ELBO(\lambda) = \mathbb{E}_{q} \left [ log p(r, z) \right ] -  \mathbb{E}_{q} \left [ log q_{\lambda}( {z|r} ) \right ] \\
= \mathbb{E}_{q} \left [ log p( {r|z} ) \right ] + \mathbb{E}_{q}  \left [ log p( {z} ) \right ] - \mathbb{E}_{q} \left [ log q_{\lambda}( {z|r} ) \right ] \scriptstyle{\text{; } log \text{ transformation of eq. \ref{eq:6}}}\\
= \mathbb{E}_{q} \left [ log p( {r|z} ) \right ]  - KL(q_{\lambda}( {z|r} ) || p ({z}))
\end{aligned}
\end{equation}

Equation \ref{eq:11} is known the VAE loss function ($L$):
\begin{equation}\label{eq:12}
L = \mathbb{E}_{q} \left [ log p( {r|z} ) \right ]  - KL(q_{\lambda}( {z|r} ) || p ({z})) \leq  log p(r)
\end{equation}

In biomolecular simulation deep neural network (DNN) encodes input data and computes $\lambda$ which approximates $q_w( {z|r}, \lambda) $. The decoder takes $p(z)$ as input and maps into original distribution $p_{w'}( {r|z} )$ . $w$ and $w'$ acts as a neural network weights. $w$ transforms the input data within the neural network hidden layers and the resulting latent variable is a combination of linear transformations that are modified by non-linear activation functions. Weights are learnable parameters during VAE training. The value of weight dictates the importance of a variable. A higher weight value corresponding an input component indicates that it will have a significant influence on the output. \\
Mathematically speaking, the reason behind VAE's popularity in biomolecular simulation community is due to the interconnectivity between \textit{loss function} and variational free energy ($F$):  
\begin{equation}\label{eq:13}
\begin{aligned}
KL(q_{\lambda}( {z|r} ) || p ({z|r})) = \int q_{\lambda}( {z|r}) log \frac{q_{\lambda}( {z|r})}{p ({z|r})} dz \\
= \int q_{\lambda}( {z|r}) log \frac{q_{\lambda}( {z|r}) p(r)}{p(r,z)} dz \\
= \int q_{\lambda}( {z|r}) log \frac{q_{\lambda}( {z|r})}{p(r,z)}dz + \int q_{\lambda}( {z|r}) log p(r) dz \\
= \int q_{\lambda}( {z|r}) log \frac{q_{\lambda}( {z|r})}{p(r,z)}dz + log p(r) \scriptstyle{\text{; as } \int q_{\lambda}( {z|r}) dz =1}\\
= - \int q_{\lambda}( {z|r}) log \frac{p(r,z)}{q_{\lambda}( {z|r)}}dz + logp(r) \\
= -F + ln p(r) 
\end{aligned}
\end{equation}

where 
\begin{equation}\label{eq:14}
\int q_{\lambda}( {z|r}) log p(r) dz = 1
\end{equation}

By comparing equation \ref{eq:9} with \ref{eq:14} we can conclude that the variational free energy (F) is equal to the $ELBO(\lambda)$. Thus minimising KL divergence is equivalent to maximising the variational free energy (F). 

\begin{figure}[h]
\hfill\includegraphics[width=10cm,height=10cm,keepaspectratio]{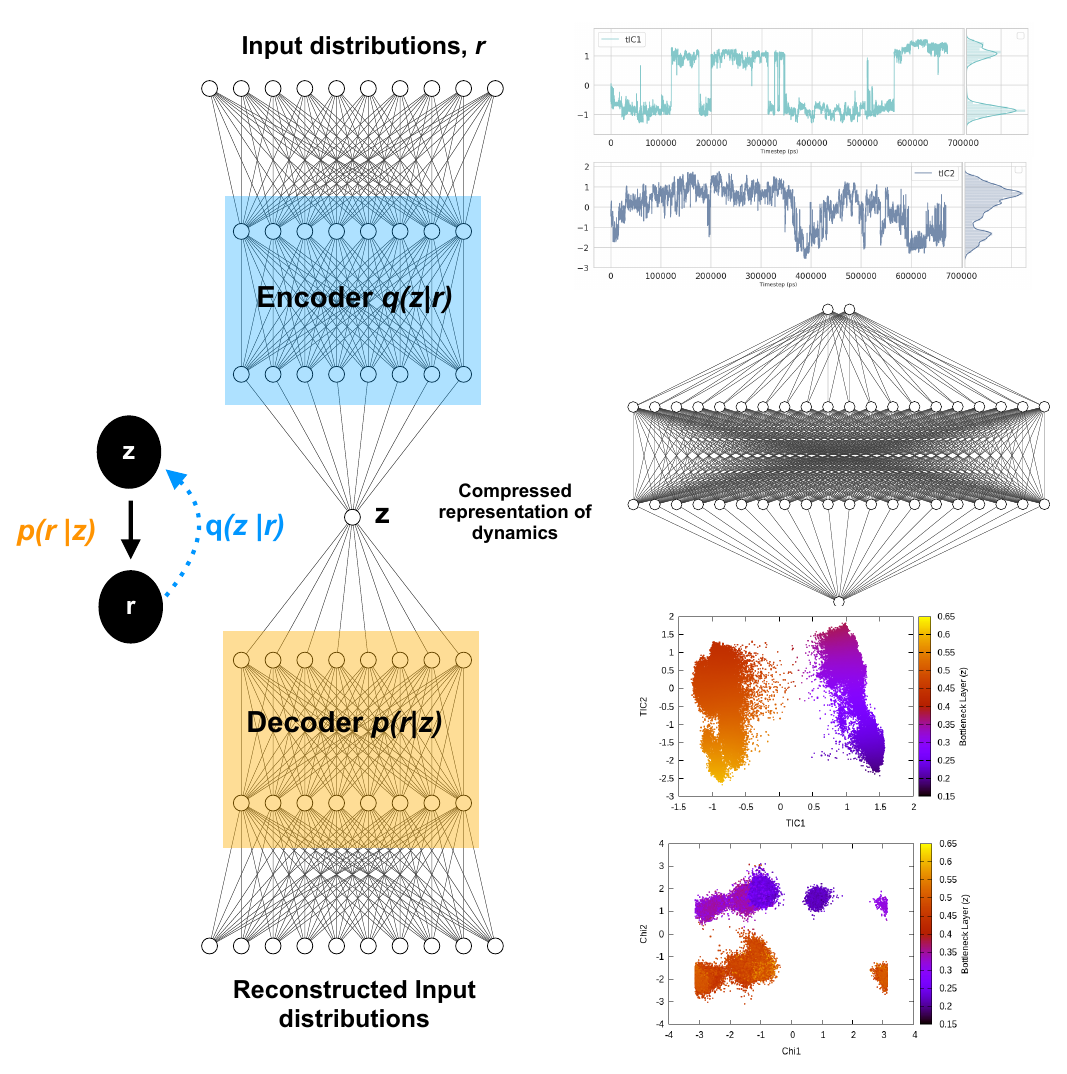}\hspace*{\fill}
\caption{Left panel showing a typical architecture of deep neural network based variational autoencoder. The input distributions $r$ can be probability distributions of geometric variables, TICs, DMD etc. Right panel shows how latent layer of variational autoencoder learned the dynamics of the system using TICs as inputs. TICs were generated using $\chi1$ and $\chi2$ angles of residues present in the flap region of plasmepsin-II (Table 1 in supplementary informations) with lag-time $1000$. Following hyper-parameters were used during the training process of VAE, no of hidden layers=$2$, no of neurons in each hidden layer=$20$, epochs=$100$, batch size=$500$, learning rate=$1e-2$.}
\label{figvae1}
\end{figure}

Figure \ref{figvae1} shows a typical architecture of VAE in context of biomolecular simulation. The encoder part of the VAE acts as a non-linear dimensionality reduction of input $\mathbf{r}$. Whereas the decoder DNN acts an input reconstruction. VAE has been primarily used a dimensionality reduction method to compress multi-variate probability distributions. The weights of the encoder layer which maps the input data onto the latent layer has been further used to drive enhanced sampling simulations such as metadynamics \cite{vdemetad}. As the latent layer of VAE represents a non-linear combination of input data hence it reduces the need of multiple CVs as inputs in metadynamics (traditionally metadynamics is limited to a maximum of two CVs). However in terms of exploration of the conformational space, metadynamics bias applied to multiple CVs\footnote{each of them as an input in VAE} (parallel-bias metadynamics) will be equivalent to using VAE's latent variable as CV in a 1D metadynamics. Several different flavours of VAEs have been developed which mainly differs in their architecture (e.g $\beta-$VAE\footnote{possible future application in biomolecular simulation to compare encoding representation difference with traditional VAE} which uses an additional hyper-parameter $\beta$ to learn disentangled latent variables by controlling the KL divergence \cite{betavae}) and types of inputs (e.g VAE-SNE which takes output of tSNE as inputs, DMD-VAE which can take multiple dynamic modes as inputs, SINDy-VAE which takes outputs from SINDy algorithm as inputs). In future, different variants of VAE can be applied on molecular dynamics simulation to capture conformational heterogeneity and drive enhanced sampling calculations. 

\textbf{Softwares \& implementations}: VAEs for post-processing high-dimensional time-series data generated from MD simulation can be implemented using popular machine learning libraries e.g PyTorch, Tensorflow, Keras etc. Recently Pande and co-workers \cite{vde}(variational dynamic encoder: \url{https://github.com/msmbuilder/vde}) as well as Tiwary and colleagues \cite{rave} (RAVE: \url{https://github.com/tiwarylab/RAVE}) applied VAE to compress high-dimensional temporal data and able to capture non-linear dynamics in context of protein folding, protein-ligand binding/unbinding etc. A collection of prebuilt VAE architectures implemented with PyTorch (\url{https://github.com/AntixK/PyTorch-VAE}) opens up the possibility to evaluate VAE variants in context of molecular dynamics simulation. 

\subsubsection{Classifiers as CVs}

\begin{figure}[h]
\hfill\includegraphics[width=12cm,height=12cm,keepaspectratio]{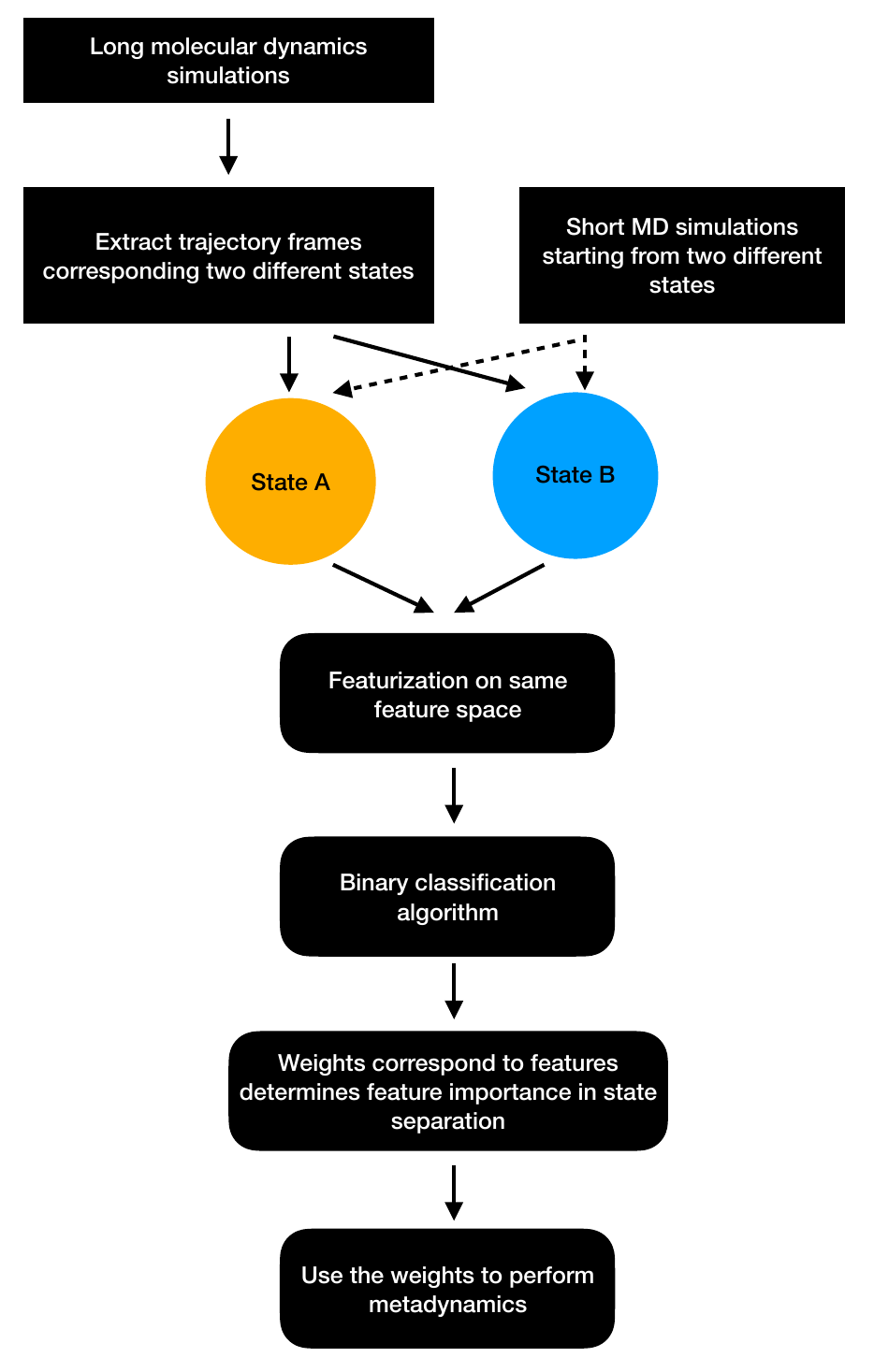}\hspace*{\fill}
\caption{Schematic representation of showing how binary classifiers can be used to separate metastable states and drive enhanced sampling calculations.}
\label{figclassifier1}
\end{figure}

Classifiers are supervised learning algorithms which uses feature vectors and feature values to map input data into specific categories \cite{classificationalgo}. In molecular dynamics simulation, the classifiers are used as CVs to recognise different metastable states\footnote{can also be used to identify features that distinguishes wild and mutant variants} based on a common set of feature values that distinguishes them (Figure \ref{figclassifier1}). Recently different classes of classifiers (e.g. support vector machine, logistic regression, artificial neural network, linear discriminant analysis etc.) have been applied to categorise different metastable states and drive enhanced sampling simulations \cite{supervisedpande,hlda}. In this article, I will discuss support vector machine (SVM) and linear discriminant analysis (LDA) and their applications in enhanced sampling. 

\begin{figure}[h]
\hfill\includegraphics[width=12cm,height=12cm,keepaspectratio]{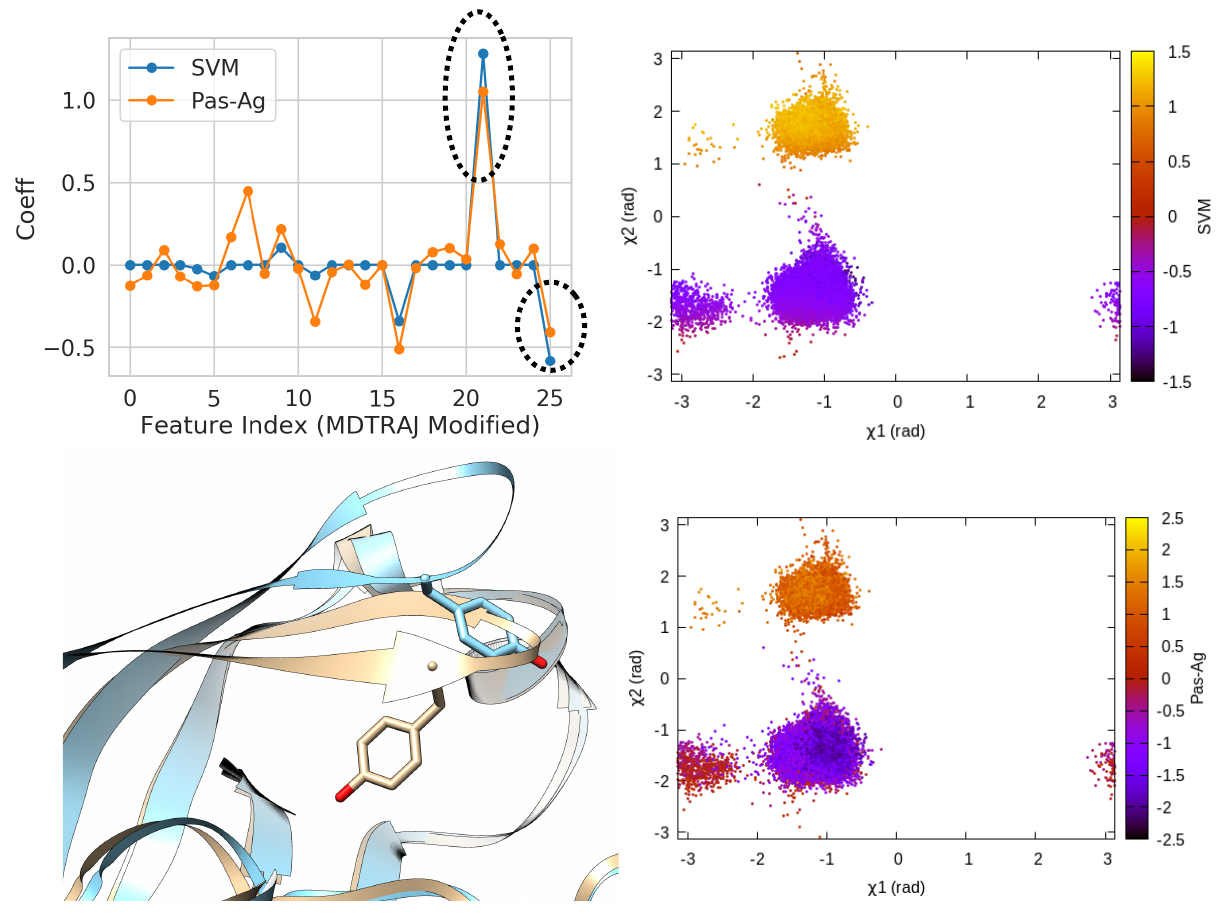}\hspace*{\fill}
\caption{Coefficients (weights) correspond to linear SVM and passive aggressive classifier\cite{passiveag} trained on $\chi1$ and $\chi2$ angles of residue 74-84 (Table 1 in Supplementary Informations) in plasmepsin-II showing the importance of $sin\chi2$ (feature $21$) and $cos\chi2$ (feature $25$) in separating two metastable basins, normal ($\chi2\sim$$-1$ rad) and flipped ($\chi2\sim$$2$ rad). 50 ns trajectories from each basins were used for training purpose using the scheme described in Figure \ref{figclassifier1}. Further $\chi1$ and $\chi2$ angles of Tyr77 projected along classifier decision boundaries demonstrated separation of metastable states. Snapshots corresponding normal (grey) and flipped (blue) states are also highlighted.}
\label{figclassifier2}
\end{figure}

\textbf{Support vector machine}:  

Let $A$ and $B$ are two metastable states which can be represented by unique points in space where each point belongs to high-dimensional feature vector $\mathbf{r}= [r_1, ......, r_N]$. The goal of SVM algorithm is to find a separating hyperplane that maximises the minimum distance between closest points (support vectors) of the two metastable states (otherwise known as \textit{classes}): 
\begin{equation}\label{eq:svm1}
d_H=\frac{\mathbf{w}^{T}\mathbf{r}+b}{\left \| w \right \|_2}
\end{equation}

where $d_H$ is the distance of a point to hyperplane, $\mathbf{w}$ is the vector of coefficients\footnote{it's direction gives us the predicted class: positive or negative}, $b$ is the intercept and $\frac{1}{{\left \| w \right \|_2}}$ is the normalisation term. The SVM algorithm then maximise the minimum distance $w*=arg_{w}max[min_{n}d_H]$. In context of biomolecular simulation, equation \ref{eq:svm1} acts a CV which can separate metastable states and drive enhanced sampling\footnote{multiple metastable states can also be classified by SVM CV followed by parallel-bias metadynamics} simulations (Figure \ref{figclassifier2}). Sultan and Pande showed how $d_H$ can be used as CV within metadynamics framework to sample conformational space of alanine dipeptide and chignolin folding \cite{supervisedpande}. SVM algorithm can be further extended for non-linear classification by using \textit{kernel} trick as explained before. 

\textbf{Linear Discriminant Analysis}:

Linear discriminant analysis (LDA) is a supervised dimensionality reduction method which finds linear combinations of features that best separates two or more metastable states \cite{fisherlda}. Imagine two metastable states with common features $\vec{r}$ have means $\vec{\mu_{A}}$ and  $\vec{\mu_{B}}$ and covariances $ \Sigma_{A}$ and $ \Sigma_{B}$. The linear combinations of features $\vec{w}\cdot\vec{r}$($\vec{w}$ is known as LDA coefficients) will have means $\vec{w}\cdot\vec{\mu_{A,B}}$ and variances $\vec{w^{T}}\Sigma_{A,B}\vec{w}$. LDA objective function separates two distributions by taking ratio between variance between the classes to variance within classes:
\begin{equation}\label{eq:ld1}
s(LDA) = \frac{(\vec{w} \cdot (\vec{\mu}_{B}-\vec{\mu}_{A}))^{2}}{\vec{w^{T}}(\Sigma_{A}+\Sigma_{B})\vec{w}}
\end{equation}
Mathematically it can shown that the maximum separation occurs when $\vec{w}\propto\frac{(\vec{\mu}_{B}-\vec{\mu}_{A})}{(\Sigma_{A}+\Sigma_{B})}$. LDA algorithm can be generalised to more than one classes (Multi-class LDA). Equation \ref{eq:ld1} acts as a CV in metadynamics to drive transition between state $A$ and $B$. Recently Parrinello and co-workers used a harmonic version of LDA (HLDA) as a CV to study folding of a small protein and protein-ligand binding/unbinding \cite{hlda,hlda2}. The framework of LDA can be patched with neural network (Deep-LDA) as well as \textit{kernel} trick in order to introduce non-linearity. Recently Deep-LDA framework has been applied in SAMPL5 host-guest systems to accurately calculate binding free energy and to understand the role of water molecule in ligand binding \cite{deepldawater}. 

In order to apply classifier algorithms to separate metastable states one first need to sample different metastable states. However, sampling of metastable states separated by high entropic barrier often requires ultra-long MD simulations or enhanced sampling simulations e.g. parallel-tempering. Further, one needs to carefully choose a set of features that can separate a set of metastable states. For complex systems, selection of such features are not trivial and often needs significant amount of trial and error. 

\textbf{Softwares}: \textit{scikit-learn's} supervised learning module incorporates both LDA and SVM algorithms for training purpose. It further integrates various other supervised learning algorithms that can be applied to classify metastable states and act as CVs in enhanced sampling simulations. 

\subsubsection{Algorithms to predict temporal evolution of CVs: proposing a challenging dataset}

\begin{figure}[!htb]
\hfill\includegraphics[width=10cm,height=10cm,keepaspectratio]{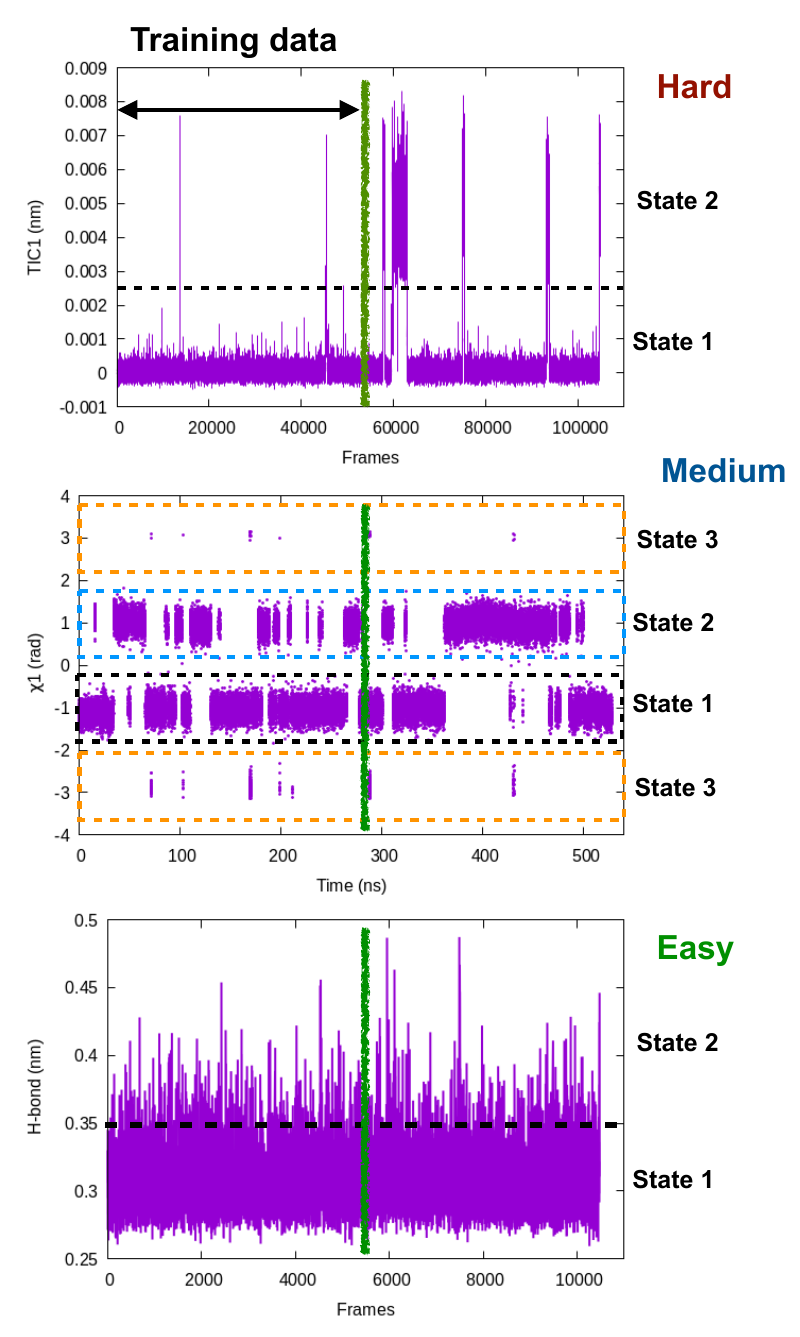}\hspace*{\fill}
\caption{Pictorial representation showing time-series fluctuation of different CVs and their hypothetical prediction difficulty using LSTM, transformers style neural network. Along H-bond distance one can see multiple recrossing between two states which makes it an $easy$ case for time-series prediction algorithms. In case of dihedral flipping along $\chi1$, training data observes several recrossing between state 1 and 2 however, there are transient sampling of rare transitions to state 3 which makes it a $medium$ level difficult task. In case of TIC1, the training data contains rare transitions between state 1 and 2 hence I believe it will be a $hard$ task for time-series prediction algorithms. One way to compare time-series projection results along a CV is compare mean first passage transition time (MFPT) as well as lifetime of different states with temporal data from long MD simulation.}
\label{figlstm}
\end{figure}

Recently neural network based time-series prediction models such as long short-term memory (LSTM) and transformers were proposed to predict rare events and extract kinetics and thermodynamics \cite{tiwarylstm,lstmrare}. These methods often work quite well on simpler systems where the training data has multiple recrossing. However in order to correctly predict kinetics from time-series one first needs to accurately predict the frequency and lifetime of rare events. In case of figure \ref{figlstm}, temporal evolution along H-bond and dihedral angle CVs showed multiple recrossing whereas evolution along TIC1 only captures transient recrossing between metastable states. Intuitively one can say predicting temporal evolution along TIC1 (\textit{rare} fluctuation) is far more challenging task when compared with H-bond or dihedral angle. No such study is currently available which compares the time-series prediction capability of neural networks in context of different CVs. In future, such study is necessary to understand the limit of neural network based time-series prediction algorithms~\cite{timeseriesforecasting} in terms of capturing temporal evolution of \textit{rare} events. 

\textbf{Softwares}: Tiwary and co-workers~\cite{tiwarylstm} used LSTM based neural network to predict temporal evolution in model systems. Their code can be accessed here: \url{https://github.com/tiwarylab/LSTM-predict-MD} .Tensorflow (\url{https://www.tensorflow.org/text/tutorials/transformer} and \url{https://www.tensorflow.org/guide/keras/rnn}) and Pytorch (\url{https://github.com/jdb78/pytorch-forecasting}) based time-series forecasting modules can be used for this purpose. Zeng and co-workers~\cite{lstmrare} used LSTM/transformers to predict temporal evolution of different CVs in alanine dipeptide (\url{https://github.com/Wendysigh/LSTM-Transformer-for-MD}) .

\section{Use of machine learning/AI in CV selection: The Hype}

\begin{figure}[h]
\hfill\includegraphics[width=10cm,height=10cm,keepaspectratio]{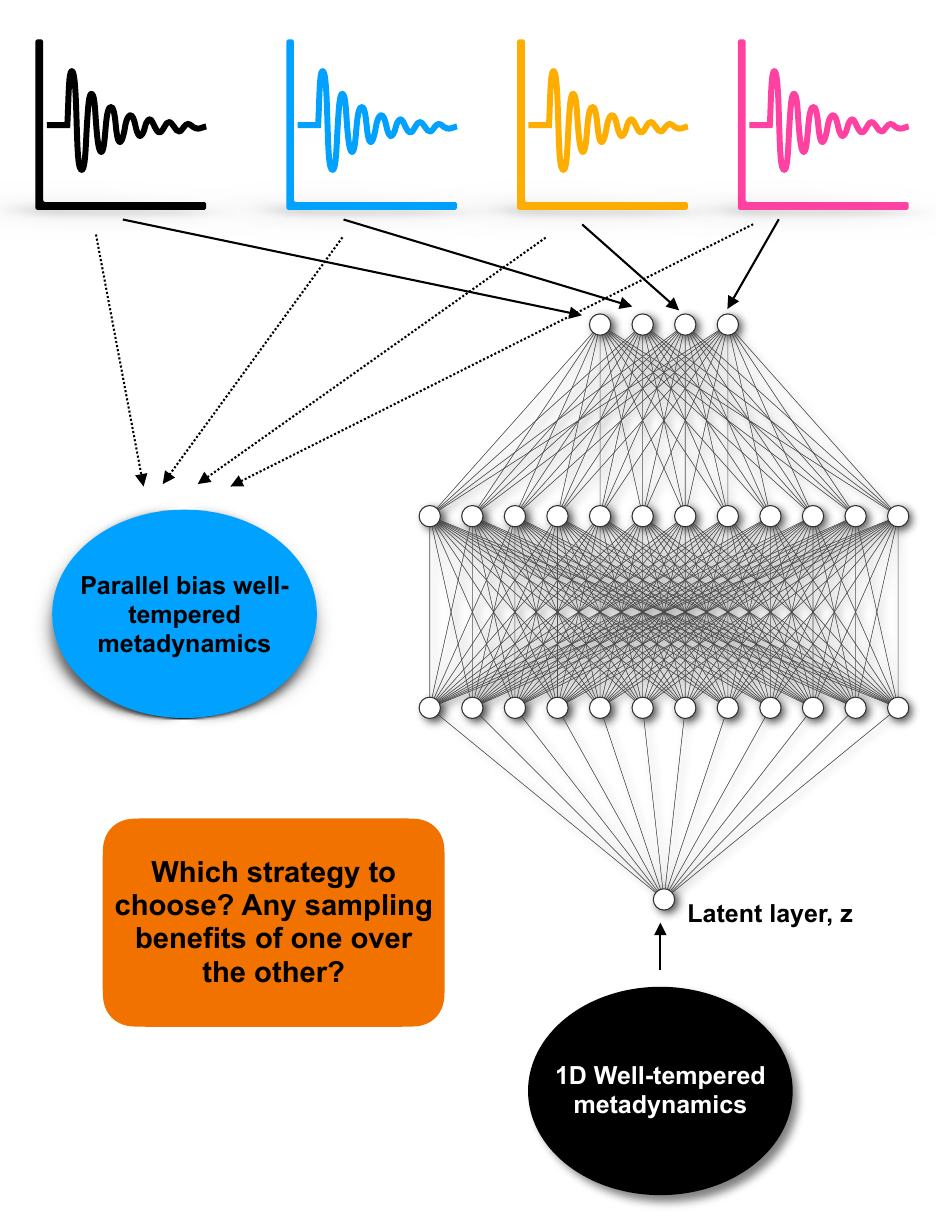}\hspace*{\fill}
\caption{Describes dilemma of choosing an enhanced sampling strategy over other: in case of four different CVs (depicted by black, blue, yellow and magenta) one can either choose to perform a parallel-bias \cite{pbmetad} variant of well-tempered metadynamics simulation or feed these four CVs within variational autoencoder and uses the latent layer of variational autoencoder as CV in 1D well-tempered metadynamics.}
\label{figpbmeta}
\end{figure}

Recently numerous papers reported different combinations of abstract order parameters to capture conformational dynamics and molecular recognition of macromolecules. In one such case neural network based LDA (Deep-LDA) approach has been applied on SAMPL5 host-guest systems to estimate binding free energy \cite{deepldawater}. Previously funnel metadynamics using simple distance CV combined with artificial restraint along binding-site solvation managed to accurately predict the binding free energy for the same host-guest systems \cite{Bhakatsampl5}. We can ask a question: was deep-LDA framework necessary for a relatively simple problem such as host-guest binding/unbinding which was previously solved using simple geometric CVs? One can argue that deep-LDA framework is more natural as it didn’t impose any artificial restraint along solvent degrees of freedom. In SAMPL5 host-guest systems, the solvation (when the binding site is occupied by long-lived water molecule) of the binding site is a slow process (Figure S2 in ref \cite{Bhakatsampl5}). In principal TICA/SGOOP (which by design capable of capturing slow degrees of freedom) could have solved such a problem. Further, deep-LDA method is a slightly fancier application of previously described work which uses binary classifiers as CVs to drive conformational sampling \cite{supervisedpande}. This is a classic example where the developers of a CV discovery method used buzzwords to solve a trivial problem without pointing out how exactly such deep learning based CV is superior compared to other geometric or abstract order parameters. \\
In another case authors used TICs/SGOOPs as inputs within a VAE and used the VAE’s latent layer as CV for enhanced sampling \cite{vdemetad,ravesgoop}. Such an approach is advantageous if an enhanced sampling method is limited to driving along one CV. Further, the eigenvalues corresponding each TIC/SGOOP can be used to filter geometric CVs which then can be used as an input in VAE. This approach can be seen as an extension of previous approach where one inputs TICs/SGOOPs directly into VAE. However a critic can ask: is there any advantage in terms of sampling if one uses VAE’s latent layer vs multiple CVs\footnote{using parallel-bias metadynamics} within metadynamics (Figure \ref{figpbmeta}) and more importantly can you call such method artificial intelligence (AI)?. 

Most of the algorithms described in the previous section either require extensive sampling or previous knowledge about metastable states in order to capture conformational changes in biomolecules. These algorithms can be best described as data driven machine learning algorithms which solves a deterministic closed set finite problems using large amount of training data. An \textit{near-term} artificial general intelligence (AGI) on other hand will learn about the dynamical system on the fly without the external supervision (e.g feature selection, choice of algorithms, tuning of hyper-parameters etc) and extract observables that can be confirmed by biophysical experiments. Development of such \textit{self-aware} AGI will require significant innovation in terms of novel algorithms and software design. Until that time, the biomolecular simulation community should refrain from using the word AI in context of identifying low-dimensional representation to approximate kinetics and thermodynamics of biomolecular systems. 

\section{Conclusions}

In recent years data-driven machine learning models impacted the field of biomolecular simulation and have been applied in context of protein folding, protein-ligand/protein binding and capturing rare conformational changes during molecular simulation. However a real challenge in comparing the power of different machine learning models is due to absence of a common dataset. We believe that the basic pancreatic trypsin inhibitor (BPTI) can act as a reasonable dataset to test different algorithms. The reasons behind that are accessibility of a) 1ms long MD simulation and b) experimental observables such as NMR order parameter, kinetics of ring flipping (rare events) and protection factors (measure of amide hydrogen-deuterium exchange). Further transient loop opening of BPTI which leads to amide hydrogen-deuterium exchange \cite{hdxbpti} and aromatic ring flipping~\cite{akkeflip,mandarflip} are the rare events that can be captured by both simulation and experiments. Transient loop opening leads to amide hydrogen-deuterium exchange which is represented by a metric called protection factor. A good machine learning model for CV discovery combined with an enhanced sampling framework should sample the rare events and predict experimental observables (rate of ring flipping and protection factors). Recent metadynamics based investigation with geometric CVs (dihedral angles) highlighted the challenge of sampling \textit{slow} ring flipping in BPTI. In order to live up to the hype, machine learning based abstract CVs should do better in sampling such rare events and predict kinetics. 

A major leap forward in combining machine learning with molecular dynamics will be prediction of multi-dimensional spatiotemporal evolution of biomolecular dynamics. Recently, transformer network has been applied to model crowd motion dynamics which managed to achieve state-of-the-art performance on commonly used pedestrian prediction datasets \cite{transformer}. In future extension of such work can be applied in context of molecular dynamics simulation to predict temporal evolution of CVs and corresponding spatial representation\footnote{such prediction will require extensive long simulations as training data}. However prediction of spatiotemporal evolution of bimolecular structure without correctly capturing the water dynamics doesn't model physical reality as water plays integral role in molecular recognition and conformational dynamics of biomolecules. We hope breakthroughs in quantum computer together with novel machine learning algorithms will make spatiotemporal prediction of solvated biomolecular system a reality, until then we have to carefully select geometric/abstract CVs to capture dynamics of biomolecules from atomistic simulations.

\begin{acknowledgement}

The computations were performed on computer resources provided by the Swedish National Infrastructure for Computing (SNIC)  at LUNARC (Lund University) and HPC2N (Ume\aa\ University). The author thanks Borna Novak for carefully reading the manuscript.

\end{acknowledgement}

\section{Conflicts of Interest}
Author declares no potential conflicts of interest.

\newpage
\begin{suppinfo}

Supplementary informations contain tables and link to Github which can be used to replicate calculations performed in this study.

\end{suppinfo}

\clearpage
\newpage

\bibliography{refrsc}

\providecommand{\latin}[1]{#1}
\makeatletter
\providecommand{\doi}
  {\begingroup\let\do\@makeother\dospecials
  \catcode`\{=1 \catcode`\}=2 \doi@aux}
\providecommand{\doi@aux}[1]{\endgroup\texttt{#1}}
\makeatother
\providecommand*\mcitethebibliography{\thebibliography}
\csname @ifundefined\endcsname{endmcitethebibliography}
  {\let\endmcitethebibliography\endthebibliography}{}
\begin{mcitethebibliography}{74}
\providecommand*\natexlab[1]{#1}
\providecommand*\mciteSetBstSublistMode[1]{}
\providecommand*\mciteSetBstMaxWidthForm[2]{}
\providecommand*\mciteBstWouldAddEndPuncttrue
  {\def\EndOfBibitem{\unskip.}}
\providecommand*\mciteBstWouldAddEndPunctfalse
  {\let\EndOfBibitem\relax}
\providecommand*\mciteSetBstMidEndSepPunct[3]{}
\providecommand*\mciteSetBstSublistLabelBeginEnd[3]{}
\providecommand*\EndOfBibitem{}
\mciteSetBstSublistMode{f}
\mciteSetBstMaxWidthForm{subitem}{(\alph{mcitesubitemcount})}
\mciteSetBstSublistLabelBeginEnd
  {\mcitemaxwidthsubitemform\space}
  {\relax}
  {\relax}

\bibitem[Hollingsworth and Dror(2018)Hollingsworth, and Dror]{drormd}
Hollingsworth,~S.~A., and Dror,~R.~O. (2018) Molecular Dynamics Simulation for
  All. \emph{Neuron} \emph{99}, 1129--1143\relax
\mciteBstWouldAddEndPuncttrue
\mciteSetBstMidEndSepPunct{\mcitedefaultmidpunct}
{\mcitedefaultendpunct}{\mcitedefaultseppunct}\relax
\EndOfBibitem
\bibitem[Fan and Lv(2010)Fan, and Lv]{featuresel}
Fan,~J., and Lv,~J. (2010) A Selective Overview of Variable Selection in High
  Dimensional Feature Space. \emph{Statistica Sinica} \emph{20}, 101--148\relax
\mciteBstWouldAddEndPuncttrue
\mciteSetBstMidEndSepPunct{\mcitedefaultmidpunct}
{\mcitedefaultendpunct}{\mcitedefaultseppunct}\relax
\EndOfBibitem
\bibitem[Bhakat(2021)]{papreview}
Bhakat,~S. (2021) Pepsin-like aspartic proteases (PAPs) as model systems for
  combining biomolecular simulation with biophysical experiments. \emph{RSC
  Adv.} \emph{11}, 11026--11047\relax
\mciteBstWouldAddEndPuncttrue
\mciteSetBstMidEndSepPunct{\mcitedefaultmidpunct}
{\mcitedefaultendpunct}{\mcitedefaultseppunct}\relax
\EndOfBibitem
\bibitem[Bhakat and S{\"o}derhjelm(2020)Bhakat, and S{\"o}derhjelm]{bhakatflap}
Bhakat,~S., and S{\"o}derhjelm,~P. (2020) Flap dynamics in pepsin-like aspartic
  proteases: a computational perspective using Plasmepsin-II and BACE-1 as
  model systems. \emph{bioRxiv} \relax
\mciteBstWouldAddEndPunctfalse
\mciteSetBstMidEndSepPunct{\mcitedefaultmidpunct}
{}{\mcitedefaultseppunct}\relax
\EndOfBibitem
\bibitem[Bussi and Laio(2020)Bussi, and Laio]{metadnaturemethods}
Bussi,~G., and Laio,~A. (2020) Using metadynamics to explore complex
  free-energy landscapes. \emph{Nature Reviews Physics} \emph{2},
  200--212\relax
\mciteBstWouldAddEndPuncttrue
\mciteSetBstMidEndSepPunct{\mcitedefaultmidpunct}
{\mcitedefaultendpunct}{\mcitedefaultseppunct}\relax
\EndOfBibitem
\bibitem[Wang \latin{et~al.}(2020)Wang, {Lamim Ribeiro}, and
  Tiwary]{tiwaryreview}
Wang,~Y., {Lamim Ribeiro},~J.~M., and Tiwary,~P. (2020) Machine learning
  approaches for analyzing and enhancing molecular dynamics simulations.
  \emph{Current Opinion in Structural Biology} \emph{61}, 139--145\relax
\mciteBstWouldAddEndPuncttrue
\mciteSetBstMidEndSepPunct{\mcitedefaultmidpunct}
{\mcitedefaultendpunct}{\mcitedefaultseppunct}\relax
\EndOfBibitem
\bibitem[Chen(2021)]{cvml}
Chen,~M. (2021) Collective variable-based enhanced sampling and machine
  learning. \emph{The European Physical Journal B} \emph{94}, 211\relax
\mciteBstWouldAddEndPuncttrue
\mciteSetBstMidEndSepPunct{\mcitedefaultmidpunct}
{\mcitedefaultendpunct}{\mcitedefaultseppunct}\relax
\EndOfBibitem
\bibitem[Valsson \latin{et~al.}(2016)Valsson, Tiwary, and
  Parrinello]{annualreviewsmetad}
Valsson,~O., Tiwary,~P., and Parrinello,~M. (2016) Enhancing Important
  Fluctuations: Rare Events and Metadynamics from a Conceptual Viewpoint.
  \emph{Annual Review of Physical Chemistry} \emph{67}, 159--184, PMID:
  26980304\relax
\mciteBstWouldAddEndPuncttrue
\mciteSetBstMidEndSepPunct{\mcitedefaultmidpunct}
{\mcitedefaultendpunct}{\mcitedefaultseppunct}\relax
\EndOfBibitem
\bibitem[Patel \latin{et~al.}(2014)Patel, Berteotti, Ronsisvalle, Rocchia, and
  Cavalli]{steeredmd}
Patel,~J.~S., Berteotti,~A., Ronsisvalle,~S., Rocchia,~W., and Cavalli,~A.
  (2014) Steered Molecular Dynamics Simulations for Studying Protein--Ligand
  Interaction in Cyclin-Dependent Kinase 5. \emph{Journal of Chemical
  Information and Modeling} \emph{54}, 470--480\relax
\mciteBstWouldAddEndPuncttrue
\mciteSetBstMidEndSepPunct{\mcitedefaultmidpunct}
{\mcitedefaultendpunct}{\mcitedefaultseppunct}\relax
\EndOfBibitem
\bibitem[You \latin{et~al.}(2019)You, Tang, and Chang]{umbrella}
You,~W., Tang,~Z., and Chang,~C.-e.~A. (2019) Potential Mean Force from
  Umbrella Sampling Simulations: What Can We Learn and What Is Missed?
  \emph{Journal of Chemical Theory and Computation} \emph{15}, 2433--2443\relax
\mciteBstWouldAddEndPuncttrue
\mciteSetBstMidEndSepPunct{\mcitedefaultmidpunct}
{\mcitedefaultendpunct}{\mcitedefaultseppunct}\relax
\EndOfBibitem
\bibitem[Dodda \latin{et~al.}(2019)Dodda, Tirado-Rives, and
  Jorgensen]{Doddameta}
Dodda,~L.~S., Tirado-Rives,~J., and Jorgensen,~W.~L. (2019) Unbinding Dynamics
  of Non-Nucleoside Inhibitors from HIV-1 Reverse Transcriptase. \emph{The
  Journal of Physical Chemistry B} \emph{123}, 1741--1748\relax
\mciteBstWouldAddEndPuncttrue
\mciteSetBstMidEndSepPunct{\mcitedefaultmidpunct}
{\mcitedefaultendpunct}{\mcitedefaultseppunct}\relax
\EndOfBibitem
\bibitem[Sittel and Stock(2018)Sittel, and Stock]{pca3}
Sittel,~F., and Stock,~G. (2018) Perspective: Identification of collective
  variables and metastable states of protein dynamics. \emph{The Journal of
  Chemical Physics} \emph{149}, 150901\relax
\mciteBstWouldAddEndPuncttrue
\mciteSetBstMidEndSepPunct{\mcitedefaultmidpunct}
{\mcitedefaultendpunct}{\mcitedefaultseppunct}\relax
\EndOfBibitem
\bibitem[Persson and Halle(2013)Persson, and Halle]{bptim1}
Persson,~F., and Halle,~B. (2013) Transient Access to the Protein Interior:
  Simulation versus NMR. \emph{Journal of the American Chemical Society}
  \emph{135}, 8735--8748\relax
\mciteBstWouldAddEndPuncttrue
\mciteSetBstMidEndSepPunct{\mcitedefaultmidpunct}
{\mcitedefaultendpunct}{\mcitedefaultseppunct}\relax
\EndOfBibitem
\bibitem[E. \latin{et~al.}(2010)E., Paul, Kresten, Stefano, O., P., A., M., K.,
  Yibing, and Willy]{desres}
E.,~S.~D., Paul,~M., Kresten,~L.-L., Stefano,~P., O.,~D.~R., P.,~E.~M.,
  A.,~B.~J., M.,~J.~J., K.,~S.~J., Yibing,~S., and Willy,~W. (2010)
  Atomic-Level Characterization of the Structural Dynamics of Proteins.
  \emph{Science} \emph{330}, 341--346\relax
\mciteBstWouldAddEndPuncttrue
\mciteSetBstMidEndSepPunct{\mcitedefaultmidpunct}
{\mcitedefaultendpunct}{\mcitedefaultseppunct}\relax
\EndOfBibitem
\bibitem[Tribello \latin{et~al.}(2014)Tribello, Bonomi, Branduardi, Camilloni,
  and Bussi]{plumed2}
Tribello,~G.~A., Bonomi,~M., Branduardi,~D., Camilloni,~C., and Bussi,~G.
  (2014) PLUMED 2: New feathers for an old bird. \emph{Computer Physics
  Communications} \emph{185}, 604--613\relax
\mciteBstWouldAddEndPuncttrue
\mciteSetBstMidEndSepPunct{\mcitedefaultmidpunct}
{\mcitedefaultendpunct}{\mcitedefaultseppunct}\relax
\EndOfBibitem
\bibitem[Roe and Cheatham(2013)Roe, and Cheatham]{cpptraj}
Roe,~D.~R., and Cheatham,~T.~E. (2013) PTRAJ and CPPTRAJ: Software for
  Processing and Analysis of Molecular Dynamics Trajectory Data. \emph{Journal
  of Chemical Theory and Computation} \emph{9}, 3084--3095\relax
\mciteBstWouldAddEndPuncttrue
\mciteSetBstMidEndSepPunct{\mcitedefaultmidpunct}
{\mcitedefaultendpunct}{\mcitedefaultseppunct}\relax
\EndOfBibitem
\bibitem[Michaud-Agrawal \latin{et~al.}(2011)Michaud-Agrawal, Denning, Woolf,
  and Beckstein]{mdanalysis}
Michaud-Agrawal,~N., Denning,~E.~J., Woolf,~T.~B., and Beckstein,~O. (2011)
  MDAnalysis: A toolkit for the analysis of molecular dynamics simulations.
  \emph{Journal of Computational Chemistry} \emph{32}, 2319--2327\relax
\mciteBstWouldAddEndPuncttrue
\mciteSetBstMidEndSepPunct{\mcitedefaultmidpunct}
{\mcitedefaultendpunct}{\mcitedefaultseppunct}\relax
\EndOfBibitem
\bibitem[McGibbon \latin{et~al.}(2015)McGibbon, Beauchamp, Harrigan, Klein,
  Swails, Hern{\'a}ndez, Schwantes, Wang, Lane, and Pande]{mdtraj}
McGibbon,~R.~T., Beauchamp,~K.~A., Harrigan,~M.~P., Klein,~C., Swails,~J.~M.,
  Hern{\'a}ndez,~C.~X., Schwantes,~C.~R., Wang,~L.-P., Lane,~T.~J., and
  Pande,~V.~S. (2015) MDTraj: A Modern Open Library for the Analysis of
  Molecular Dynamics Trajectories. \emph{Biophysical Journal} \emph{109},
  1528--1532\relax
\mciteBstWouldAddEndPuncttrue
\mciteSetBstMidEndSepPunct{\mcitedefaultmidpunct}
{\mcitedefaultendpunct}{\mcitedefaultseppunct}\relax
\EndOfBibitem
\bibitem[Harrigan \latin{et~al.}(2017)Harrigan, Sultan, Hern{\'a}ndez, Husic,
  Eastman, Schwantes, Beauchamp, McGibbon, and Pande]{msmbuilder}
Harrigan,~M.~P., Sultan,~M.~M., Hern{\'a}ndez,~C.~X., Husic,~B.~E.,
  Eastman,~P., Schwantes,~C.~R., Beauchamp,~K.~A., McGibbon,~R.~T., and
  Pande,~V.~S. (2017) MSMBuilder: Statistical Models for Biomolecular Dynamics.
  \emph{Biophysical Journal} \emph{112}, 10--15\relax
\mciteBstWouldAddEndPuncttrue
\mciteSetBstMidEndSepPunct{\mcitedefaultmidpunct}
{\mcitedefaultendpunct}{\mcitedefaultseppunct}\relax
\EndOfBibitem
\bibitem[Scherer \latin{et~al.}(2015)Scherer, Trendelkamp-Schroer, Paul,
  P{\'e}rez-Hern{\'a}ndez, Hoffmann, Plattner, Wehmeyer, Prinz, and
  No{\'e}]{pyemma}
Scherer,~M.~K., Trendelkamp-Schroer,~B., Paul,~F., P{\'e}rez-Hern{\'a}ndez,~G.,
  Hoffmann,~M., Plattner,~N., Wehmeyer,~C., Prinz,~J.-H., and No{\'e},~F.
  (2015) PyEMMA 2: A Software Package for Estimation, Validation, and Analysis
  of Markov Models. \emph{Journal of Chemical Theory and Computation}
  \emph{11}, 5525--5542\relax
\mciteBstWouldAddEndPuncttrue
\mciteSetBstMidEndSepPunct{\mcitedefaultmidpunct}
{\mcitedefaultendpunct}{\mcitedefaultseppunct}\relax
\EndOfBibitem
\bibitem[Molgedey and Schuster(1994)Molgedey, and Schuster]{schustertica}
Molgedey,~L., and Schuster,~H.~G. (1994) Separation of a mixture of independent
  signals using time delayed correlations. \emph{Phys. Rev. Lett.} \emph{72},
  3634--3637\relax
\mciteBstWouldAddEndPuncttrue
\mciteSetBstMidEndSepPunct{\mcitedefaultmidpunct}
{\mcitedefaultendpunct}{\mcitedefaultseppunct}\relax
\EndOfBibitem
\bibitem[P{\'e}rez-Hern{\'a}ndez \latin{et~al.}(2013)P{\'e}rez-Hern{\'a}ndez,
  Paul, Giorgino, De~Fabritiis, and No{\'e}]{noetica}
P{\'e}rez-Hern{\'a}ndez,~G., Paul,~F., Giorgino,~T., De~Fabritiis,~G., and
  No{\'e},~F. (2013) Identification of slow molecular order parameters for
  Markov model construction. \emph{The Journal of Chemical Physics} \emph{139},
  015102\relax
\mciteBstWouldAddEndPuncttrue
\mciteSetBstMidEndSepPunct{\mcitedefaultmidpunct}
{\mcitedefaultendpunct}{\mcitedefaultseppunct}\relax
\EndOfBibitem
\bibitem[Schwantes and Pande(2013)Schwantes, and Pande]{pandetica}
Schwantes,~C.~R., and Pande,~V.~S. (2013) Improvements in Markov State Model
  Construction Reveal Many Non-Native Interactions in the Folding of NTL9.
  \emph{Journal of Chemical Theory and Computation} \emph{9}, 2000--2009, PMID:
  23750122\relax
\mciteBstWouldAddEndPuncttrue
\mciteSetBstMidEndSepPunct{\mcitedefaultmidpunct}
{\mcitedefaultendpunct}{\mcitedefaultseppunct}\relax
\EndOfBibitem
\bibitem[M.~Sultan and Pande(2017)M.~Sultan, and Pande]{ticametad}
M.~Sultan,~M., and Pande,~V.~S. (2017) tICA-Metadynamics: Accelerating
  Metadynamics by Using Kinetically Selected Collective Variables.
  \emph{Journal of Chemical Theory and Computation} \emph{13}, 2440--2447,
  PMID: 28383914\relax
\mciteBstWouldAddEndPuncttrue
\mciteSetBstMidEndSepPunct{\mcitedefaultmidpunct}
{\mcitedefaultendpunct}{\mcitedefaultseppunct}\relax
\EndOfBibitem
\bibitem[Schultze and Grubm{\"u}ller(2021)Schultze, and
  Grubm{\"u}ller]{schultze}
Schultze,~S., and Grubm{\"u}ller,~H. (2021) Time-Lagged Independent Component
  Analysis of Random Walks and Protein Dynamics. \emph{Journal of Chemical
  Theory and Computation} \emph{17}, 5766--5776\relax
\mciteBstWouldAddEndPuncttrue
\mciteSetBstMidEndSepPunct{\mcitedefaultmidpunct}
{\mcitedefaultendpunct}{\mcitedefaultseppunct}\relax
\EndOfBibitem
\bibitem[Blaschke \latin{et~al.}(2006)Blaschke, Berkes, and Wiskott]{sfavsica}
Blaschke,~T., Berkes,~P., and Wiskott,~L. (2006) {What Is the Relation Between
  Slow Feature Analysis and Independent Component Analysis?} \emph{Neural
  Computation} \emph{18}, 2495--2508\relax
\mciteBstWouldAddEndPuncttrue
\mciteSetBstMidEndSepPunct{\mcitedefaultmidpunct}
{\mcitedefaultendpunct}{\mcitedefaultseppunct}\relax
\EndOfBibitem
\bibitem[Blaschke and Wiskott(2004)Blaschke, and Wiskott]{sfaandica}
Blaschke,~T., and Wiskott,~L. Independent Slow Feature Analysis and Nonlinear
  Blind Source Separation. Independent Component Analysis and Blind Signal
  Separation. Berlin, Heidelberg, 2004; pp 742--749\relax
\mciteBstWouldAddEndPuncttrue
\mciteSetBstMidEndSepPunct{\mcitedefaultmidpunct}
{\mcitedefaultendpunct}{\mcitedefaultseppunct}\relax
\EndOfBibitem
\bibitem[Persson and Halle(2015)Persson, and Halle]{hdxbpti}
Persson,~F., and Halle,~B. (2015) How amide hydrogens exchange in native
  proteins. \emph{Proceedings of the National Academy of Sciences} \emph{112},
  10383--10388\relax
\mciteBstWouldAddEndPuncttrue
\mciteSetBstMidEndSepPunct{\mcitedefaultmidpunct}
{\mcitedefaultendpunct}{\mcitedefaultseppunct}\relax
\EndOfBibitem
\bibitem[Hoffmann \latin{et~al.}(2021)Hoffmann, Scherer, Hempel, Mardt,
  de~Silva, Husic, Klus, Wu, Kutz, Brunton, and No{\'e}]{deeptime}
Hoffmann,~M., Scherer,~M., Hempel,~T., Mardt,~A., de~Silva,~B., Husic,~B.~E.,
  Klus,~S., Wu,~H., Kutz,~N., Brunton,~S.~L., and No{\'e},~F. Deeptime: a
  Python library for machine learning dynamical models from time series data.
  2021\relax
\mciteBstWouldAddEndPuncttrue
\mciteSetBstMidEndSepPunct{\mcitedefaultmidpunct}
{\mcitedefaultendpunct}{\mcitedefaultseppunct}\relax
\EndOfBibitem
\bibitem[Schwantes and Pande(2015)Schwantes, and Pande]{kerneltica}
Schwantes,~C.~R., and Pande,~V.~S. (2015) Modeling Molecular Kinetics with tICA
  and the Kernel Trick. \emph{Journal of Chemical Theory and Computation}
  \emph{11}, 600--608\relax
\mciteBstWouldAddEndPuncttrue
\mciteSetBstMidEndSepPunct{\mcitedefaultmidpunct}
{\mcitedefaultendpunct}{\mcitedefaultseppunct}\relax
\EndOfBibitem
\bibitem[Harrigan and Pande(2017)Harrigan, and Pande]{landmarkkernel}
Harrigan,~M.~P., and Pande,~V.~S. (2017) Landmark Kernel tICA for
  Conformational Dynamics. \emph{bioRxiv} \relax
\mciteBstWouldAddEndPunctfalse
\mciteSetBstMidEndSepPunct{\mcitedefaultmidpunct}
{}{\mcitedefaultseppunct}\relax
\EndOfBibitem
\bibitem[Ross \latin{et~al.}(2018)Ross, Nizami, Glenister, Sheik~Amamuddy,
  Atilgan, Atilgan, and Tastan~Bishop]{modetask}
Ross,~C., Nizami,~B., Glenister,~M., Sheik~Amamuddy,~O., Atilgan,~A.~R.,
  Atilgan,~C., and Tastan~Bishop,~{\"O}. (2018) {MODE-TASK: large-scale protein
  motion tools}. \emph{Bioinformatics} \emph{34}, 3759--3763\relax
\mciteBstWouldAddEndPuncttrue
\mciteSetBstMidEndSepPunct{\mcitedefaultmidpunct}
{\mcitedefaultendpunct}{\mcitedefaultseppunct}\relax
\EndOfBibitem
\bibitem[No{\'e} and Clementi(2015)No{\'e}, and Clementi]{Noeclementi}
No{\'e},~F., and Clementi,~C. (2015) Kinetic Distance and Kinetic Maps from
  Molecular Dynamics Simulation. \emph{Journal of Chemical Theory and
  Computation} \emph{11}, 5002--5011\relax
\mciteBstWouldAddEndPuncttrue
\mciteSetBstMidEndSepPunct{\mcitedefaultmidpunct}
{\mcitedefaultendpunct}{\mcitedefaultseppunct}\relax
\EndOfBibitem
\bibitem[Nadler \latin{et~al.}(2006)Nadler, Lafon, Kevrekidis, and
  Coifman]{dmapneuralips}
Nadler,~B., Lafon,~S., Kevrekidis,~I., and Coifman,~R. Diffusion Maps, Spectral
  Clustering and Eigenfunctions of Fokker-Planck Operators. Advances in Neural
  Information Processing Systems. 2006\relax
\mciteBstWouldAddEndPuncttrue
\mciteSetBstMidEndSepPunct{\mcitedefaultmidpunct}
{\mcitedefaultendpunct}{\mcitedefaultseppunct}\relax
\EndOfBibitem
\bibitem[Zheng \latin{et~al.}(2013)Zheng, Rohrdanz, and
  Clementi]{diffusionmap1}
Zheng,~W., Rohrdanz,~M.~A., and Clementi,~C. (2013) Rapid Exploration of
  Configuration Space with Diffusion-Map-Directed Molecular Dynamics. \emph{The
  Journal of Physical Chemistry B} \emph{117}, 12769--12776\relax
\mciteBstWouldAddEndPuncttrue
\mciteSetBstMidEndSepPunct{\mcitedefaultmidpunct}
{\mcitedefaultendpunct}{\mcitedefaultseppunct}\relax
\EndOfBibitem
\bibitem[Ferguson \latin{et~al.}(2010)Ferguson, Panagiotopoulos, Debenedetti,
  and Kevrekidis]{diffusionmap2}
Ferguson,~A.~L., Panagiotopoulos,~A.~Z., Debenedetti,~P.~G., and
  Kevrekidis,~I.~G. (2010) Systematic determination of order parameters for
  chain dynamics using diffusion maps. \emph{Proceedings of the National
  Academy of Sciences} \emph{107}, 13597--13602\relax
\mciteBstWouldAddEndPuncttrue
\mciteSetBstMidEndSepPunct{\mcitedefaultmidpunct}
{\mcitedefaultendpunct}{\mcitedefaultseppunct}\relax
\EndOfBibitem
\bibitem[Ferguson \latin{et~al.}(2011)Ferguson, Panagiotopoulos, Kevrekidis,
  and Debenedetti]{diffusionmap3}
Ferguson,~A.~L., Panagiotopoulos,~A.~Z., Kevrekidis,~I.~G., and
  Debenedetti,~P.~G. (2011) Nonlinear dimensionality reduction in molecular
  simulation: The diffusion map approach. \emph{Chemical Physics Letters}
  \emph{509}, 1--11\relax
\mciteBstWouldAddEndPuncttrue
\mciteSetBstMidEndSepPunct{\mcitedefaultmidpunct}
{\mcitedefaultendpunct}{\mcitedefaultseppunct}\relax
\EndOfBibitem
\bibitem[Tsai \latin{et~al.}(2021)Tsai, Smith, and Tiwary]{sgoopd}
Tsai,~S.-T., Smith,~Z., and Tiwary,~P. (2021) SGOOP-d: Estimating Kinetic
  Distances and Reaction Coordinate Dimensionality for Rare Event Systems from
  Biased/Unbiased Simulations. \emph{Journal of Chemical Theory and
  Computation} \emph{17}, 6757--6765\relax
\mciteBstWouldAddEndPuncttrue
\mciteSetBstMidEndSepPunct{\mcitedefaultmidpunct}
{\mcitedefaultendpunct}{\mcitedefaultseppunct}\relax
\EndOfBibitem
\bibitem[van~der Maaten and Hinton(2008)van~der Maaten, and Hinton]{JMLRtsne}
van~der Maaten,~L., and Hinton,~G. (2008) Visualizing Data using t-SNE.
  \emph{Journal of Machine Learning Research} \emph{9}, 2579--2605\relax
\mciteBstWouldAddEndPuncttrue
\mciteSetBstMidEndSepPunct{\mcitedefaultmidpunct}
{\mcitedefaultendpunct}{\mcitedefaultseppunct}\relax
\EndOfBibitem
\bibitem[Wattenberg \latin{et~al.}(2016)Wattenberg, Vi{\'e}gas, and
  Johnson]{tsnedistill}
Wattenberg,~M., Vi{\'e}gas,~F., and Johnson,~I. (2016) How to Use t-SNE
  Effectively. \emph{Distill} \relax
\mciteBstWouldAddEndPunctfalse
\mciteSetBstMidEndSepPunct{\mcitedefaultmidpunct}
{}{\mcitedefaultseppunct}\relax
\EndOfBibitem
\bibitem[Spiwok and K{\v r}{\'\i}{\v z}(2020)Spiwok, and K{\v r}{\'\i}{\v
  z}]{timelagtsne}
Spiwok,~V., and K{\v r}{\'\i}{\v z},~P. (2020) Time-Lagged t-Distributed
  Stochastic Neighbor Embedding (t-SNE) of Molecular Simulation Trajectories.
  \emph{Frontiers in Molecular Biosciences} \emph{7}, 132\relax
\mciteBstWouldAddEndPuncttrue
\mciteSetBstMidEndSepPunct{\mcitedefaultmidpunct}
{\mcitedefaultendpunct}{\mcitedefaultseppunct}\relax
\EndOfBibitem
\bibitem[Chari \latin{et~al.}(2021)Chari, Banerjee, and Pachter]{liortsne}
Chari,~T., Banerjee,~J., and Pachter,~L. (2021) The Specious Art of Single-Cell
  Genomics. \emph{bioRxiv} \relax
\mciteBstWouldAddEndPunctfalse
\mciteSetBstMidEndSepPunct{\mcitedefaultmidpunct}
{}{\mcitedefaultseppunct}\relax
\EndOfBibitem
\bibitem[Tiwary and Berne(2016)Tiwary, and Berne]{sgoop}
Tiwary,~P., and Berne,~B.~J. (2016) Spectral gap optimization of order
  parameters for sampling complex molecular systems. \emph{Proceedings of the
  National Academy of Sciences} \emph{113}, 2839--2844\relax
\mciteBstWouldAddEndPuncttrue
\mciteSetBstMidEndSepPunct{\mcitedefaultmidpunct}
{\mcitedefaultendpunct}{\mcitedefaultseppunct}\relax
\EndOfBibitem
\bibitem[Ghojogh \latin{et~al.}(2020)Ghojogh, Ghodsi, Karray, and
  Crowley]{mdsisomap}
Ghojogh,~B., Ghodsi,~A., Karray,~F., and Crowley,~M. Multidimensional Scaling,
  Sammon Mapping, and Isomap: Tutorial and Survey. 2020\relax
\mciteBstWouldAddEndPuncttrue
\mciteSetBstMidEndSepPunct{\mcitedefaultmidpunct}
{\mcitedefaultendpunct}{\mcitedefaultseppunct}\relax
\EndOfBibitem
\bibitem[Tu \latin{et~al.}(2014)Tu, Rowley, Luchtenburg, Brunton, and
  Kutz]{dmd}
Tu,~J.~H., Rowley,~C.~W., Luchtenburg,~D.~M., Brunton,~S.~L., and Kutz,~J.~N.
  (2014) On dynamic mode decomposition: Theory and applications. \emph{Journal
  of Computational Dynamics} \emph{1}, 391--421\relax
\mciteBstWouldAddEndPuncttrue
\mciteSetBstMidEndSepPunct{\mcitedefaultmidpunct}
{\mcitedefaultendpunct}{\mcitedefaultseppunct}\relax
\EndOfBibitem
\bibitem[Klus \latin{et~al.}(2018)Klus, N{\"u}ske, Koltai, Wu, Kevrekidis,
  Sch{\"u}tte, and No{\'e}]{datadriven}
Klus,~S., N{\"u}ske,~F., Koltai,~P., Wu,~H., Kevrekidis,~I., Sch{\"u}tte,~C.,
  and No{\'e},~F. (2018) Data-Driven Model Reduction and Transfer Operator
  Approximation. \emph{Journal of Nonlinear Science} \emph{28}, 985--1010\relax
\mciteBstWouldAddEndPuncttrue
\mciteSetBstMidEndSepPunct{\mcitedefaultmidpunct}
{\mcitedefaultendpunct}{\mcitedefaultseppunct}\relax
\EndOfBibitem
\bibitem[McInnes \latin{et~al.}(2020)McInnes, Healy, and Melville]{umap}
McInnes,~L., Healy,~J., and Melville,~J. UMAP: Uniform Manifold Approximation
  and Projection for Dimension Reduction. 2020\relax
\mciteBstWouldAddEndPuncttrue
\mciteSetBstMidEndSepPunct{\mcitedefaultmidpunct}
{\mcitedefaultendpunct}{\mcitedefaultseppunct}\relax
\EndOfBibitem
\bibitem[Trozzi \latin{et~al.}(2021)Trozzi, Wang, and Tao]{umapmd}
Trozzi,~F., Wang,~X., and Tao,~P. (2021) UMAP as a Dimensionality Reduction
  Tool for Molecular Dynamics Simulations of Biomacromolecules: A Comparison
  Study. \emph{The Journal of Physical Chemistry B} \emph{125}, 5022--5034,
  PMID: 33973773\relax
\mciteBstWouldAddEndPuncttrue
\mciteSetBstMidEndSepPunct{\mcitedefaultmidpunct}
{\mcitedefaultendpunct}{\mcitedefaultseppunct}\relax
\EndOfBibitem
\bibitem[Tian and Tao(2020)Tian, and Tao]{ivis}
Tian,~H., and Tao,~P. (2020) ivis Dimensionality Reduction Framework for
  Biomacromolecular Simulations. \emph{Journal of Chemical Information and
  Modeling} \emph{60}, 4569--4581\relax
\mciteBstWouldAddEndPuncttrue
\mciteSetBstMidEndSepPunct{\mcitedefaultmidpunct}
{\mcitedefaultendpunct}{\mcitedefaultseppunct}\relax
\EndOfBibitem
\bibitem[Brunton \latin{et~al.}(2016)Brunton, Proctor, and Kutz]{sindy}
Brunton,~S.~L., Proctor,~J.~L., and Kutz,~J.~N. (2016) Discovering governing
  equations from data by sparse identification of nonlinear dynamical systems.
  \emph{Proceedings of the National Academy of Sciences} \emph{113},
  3932--3937\relax
\mciteBstWouldAddEndPuncttrue
\mciteSetBstMidEndSepPunct{\mcitedefaultmidpunct}
{\mcitedefaultendpunct}{\mcitedefaultseppunct}\relax
\EndOfBibitem
\bibitem[Glielmo \latin{et~al.}(2021)Glielmo, Husic, Rodriguez, Clementi,
  No{\'e}, and Laio]{noe-chemicalreviews}
Glielmo,~A., Husic,~B.~E., Rodriguez,~A., Clementi,~C., No{\'e},~F., and
  Laio,~A. (2021) Unsupervised Learning Methods for Molecular Simulation Data.
  \emph{Chemical Reviews} \emph{121}, 9722--9758, PMID: 33945269\relax
\mciteBstWouldAddEndPuncttrue
\mciteSetBstMidEndSepPunct{\mcitedefaultmidpunct}
{\mcitedefaultendpunct}{\mcitedefaultseppunct}\relax
\EndOfBibitem
\bibitem[Kingma and Welling(2019)Kingma, and Welling]{vaewelling}
Kingma,~D.~P., and Welling,~M. (2019) An Introduction to Variational
  Autoencoders. \emph{Foundations and Trends{\textregistered} in Machine
  Learning} \emph{12}, 307--392\relax
\mciteBstWouldAddEndPuncttrue
\mciteSetBstMidEndSepPunct{\mcitedefaultmidpunct}
{\mcitedefaultendpunct}{\mcitedefaultseppunct}\relax
\EndOfBibitem
\bibitem[Kingma and Welling(2014)Kingma, and Welling]{autoencoding}
Kingma,~D.~P., and Welling,~M. Auto-Encoding Variational Bayes. 2014\relax
\mciteBstWouldAddEndPuncttrue
\mciteSetBstMidEndSepPunct{\mcitedefaultmidpunct}
{\mcitedefaultendpunct}{\mcitedefaultseppunct}\relax
\EndOfBibitem
\bibitem[Sultan \latin{et~al.}(2018)Sultan, Wayment-Steele, and
  Pande]{vdemetad}
Sultan,~M.~M., Wayment-Steele,~H.~K., and Pande,~V.~S. (2018) Transferable
  Neural Networks for Enhanced Sampling of Protein Dynamics. \emph{Journal of
  Chemical Theory and Computation} \emph{14}, 1887--1894\relax
\mciteBstWouldAddEndPuncttrue
\mciteSetBstMidEndSepPunct{\mcitedefaultmidpunct}
{\mcitedefaultendpunct}{\mcitedefaultseppunct}\relax
\EndOfBibitem
\bibitem[Burgess \latin{et~al.}(2018)Burgess, Higgins, Pal, Matthey, Watters,
  Desjardins, and Lerchner]{betavae}
Burgess,~C.~P., Higgins,~I., Pal,~A., Matthey,~L., Watters,~N., Desjardins,~G.,
  and Lerchner,~A. Understanding disentangling in $\beta$-VAE. 2018\relax
\mciteBstWouldAddEndPuncttrue
\mciteSetBstMidEndSepPunct{\mcitedefaultmidpunct}
{\mcitedefaultendpunct}{\mcitedefaultseppunct}\relax
\EndOfBibitem
\bibitem[Hern\'andez \latin{et~al.}(2018)Hern\'andez, Wayment-Steele, Sultan,
  Husic, and Pande]{vde}
Hern\'andez,~C.~X., Wayment-Steele,~H.~K., Sultan,~M.~M., Husic,~B.~E., and
  Pande,~V.~S. (2018) Variational encoding of complex dynamics. \emph{Phys.
  Rev. E} \emph{97}, 062412\relax
\mciteBstWouldAddEndPuncttrue
\mciteSetBstMidEndSepPunct{\mcitedefaultmidpunct}
{\mcitedefaultendpunct}{\mcitedefaultseppunct}\relax
\EndOfBibitem
\bibitem[Ribeiro \latin{et~al.}(2018)Ribeiro, Bravo, Wang, and Tiwary]{rave}
Ribeiro,~J. M.~L., Bravo,~P., Wang,~Y., and Tiwary,~P. (2018) Reweighted
  autoencoded variational Bayes for enhanced sampling (RAVE). \emph{The Journal
  of Chemical Physics} \emph{149}, 072301\relax
\mciteBstWouldAddEndPuncttrue
\mciteSetBstMidEndSepPunct{\mcitedefaultmidpunct}
{\mcitedefaultendpunct}{\mcitedefaultseppunct}\relax
\EndOfBibitem
\bibitem[Mohanty \latin{et~al.}(2013)Mohanty, John, Manmatha, and
  Rath]{classificationalgo}
Mohanty,~N., John,~A. L.-S., Manmatha,~R., and Rath,~T. In \emph{Handbook of
  Statistics}; Rao,~C., and Govindaraju,~V., Eds.; Handbook of Statistics;
  Elsevier, 2013; Vol.~31; pp 249--267\relax
\mciteBstWouldAddEndPuncttrue
\mciteSetBstMidEndSepPunct{\mcitedefaultmidpunct}
{\mcitedefaultendpunct}{\mcitedefaultseppunct}\relax
\EndOfBibitem
\bibitem[Sultan and Pande(2018)Sultan, and Pande]{supervisedpande}
Sultan,~M.~M., and Pande,~V.~S. (2018) Automated design of collective variables
  using supervised machine learning. \emph{The Journal of Chemical Physics}
  \emph{149}, 094106\relax
\mciteBstWouldAddEndPuncttrue
\mciteSetBstMidEndSepPunct{\mcitedefaultmidpunct}
{\mcitedefaultendpunct}{\mcitedefaultseppunct}\relax
\EndOfBibitem
\bibitem[Mendels \latin{et~al.}(2018)Mendels, Piccini, Brotzakis, Yang, and
  Parrinello]{hlda}
Mendels,~D., Piccini,~G., Brotzakis,~Z.~F., Yang,~Y.~I., and Parrinello,~M.
  (2018) Folding a small protein using harmonic linear discriminant analysis.
  \emph{The Journal of Chemical Physics} \emph{149}, 194113\relax
\mciteBstWouldAddEndPuncttrue
\mciteSetBstMidEndSepPunct{\mcitedefaultmidpunct}
{\mcitedefaultendpunct}{\mcitedefaultseppunct}\relax
\EndOfBibitem
\bibitem[Crammer \latin{et~al.}(2006)Crammer, Dekel, Keshet, Shalev-Shwartz,
  and Singer]{passiveag}
Crammer,~K., Dekel,~O., Keshet,~J., Shalev-Shwartz,~S., and Singer,~Y. (2006)
  Online Passive-Aggressive Algorithms. \emph{Journal of Machine Learning
  Research} \emph{7}, 551--585\relax
\mciteBstWouldAddEndPuncttrue
\mciteSetBstMidEndSepPunct{\mcitedefaultmidpunct}
{\mcitedefaultendpunct}{\mcitedefaultseppunct}\relax
\EndOfBibitem
\bibitem[Fisher(1936)]{fisherlda}
Fisher,~R.~A. (1936) The use of multiple measurements in taxonomic problems.
  \emph{Annals of Eugenics} \emph{7}, 179--188\relax
\mciteBstWouldAddEndPuncttrue
\mciteSetBstMidEndSepPunct{\mcitedefaultmidpunct}
{\mcitedefaultendpunct}{\mcitedefaultseppunct}\relax
\EndOfBibitem
\bibitem[Capelli \latin{et~al.}(2019)Capelli, Bochicchio, Piccini, Casasnovas,
  Carloni, and Parrinello]{hlda2}
Capelli,~R., Bochicchio,~A., Piccini,~G., Casasnovas,~R., Carloni,~P., and
  Parrinello,~M. (2019) Chasing the Full Free Energy Landscape of
  Neuroreceptor/Ligand Unbinding by Metadynamics Simulations. \emph{Journal of
  Chemical Theory and Computation} \emph{15}, 3354--3361\relax
\mciteBstWouldAddEndPuncttrue
\mciteSetBstMidEndSepPunct{\mcitedefaultmidpunct}
{\mcitedefaultendpunct}{\mcitedefaultseppunct}\relax
\EndOfBibitem
\bibitem[Rizzi \latin{et~al.}(2021)Rizzi, Bonati, Ansari, and
  Parrinello]{deepldawater}
Rizzi,~V., Bonati,~L., Ansari,~N., and Parrinello,~M. (2021) The role of water
  in host-guest interaction. \emph{Nature Communications} \emph{12}, 93\relax
\mciteBstWouldAddEndPuncttrue
\mciteSetBstMidEndSepPunct{\mcitedefaultmidpunct}
{\mcitedefaultendpunct}{\mcitedefaultseppunct}\relax
\EndOfBibitem
\bibitem[Tsai \latin{et~al.}(2020)Tsai, Kuo, and Tiwary]{tiwarylstm}
Tsai,~S.-T., Kuo,~E.-J., and Tiwary,~P. (2020) Learning molecular dynamics with
  simple language model built upon long short-term memory neural network.
  \emph{Nature Communications} \emph{11}, 5115\relax
\mciteBstWouldAddEndPuncttrue
\mciteSetBstMidEndSepPunct{\mcitedefaultmidpunct}
{\mcitedefaultendpunct}{\mcitedefaultseppunct}\relax
\EndOfBibitem
\bibitem[Zeng \latin{et~al.}(2021)Zeng, Cao, Huang, and Yao]{lstmrare}
Zeng,~W., Cao,~S., Huang,~X., and Yao,~Y. A Note on Learning Rare Events in
  Molecular Dynamics using LSTM and Transformer. 2021\relax
\mciteBstWouldAddEndPuncttrue
\mciteSetBstMidEndSepPunct{\mcitedefaultmidpunct}
{\mcitedefaultendpunct}{\mcitedefaultseppunct}\relax
\EndOfBibitem
\bibitem[Lim and Zohren(2021)Lim, and Zohren]{timeseriesforecasting}
Lim,~B., and Zohren,~S. (2021) Time-series forecasting with deep learning: a
  survey. \emph{Philosophical Transactions of the Royal Society A:
  Mathematical, Physical and Engineering Sciences} \emph{379}, 20200209\relax
\mciteBstWouldAddEndPuncttrue
\mciteSetBstMidEndSepPunct{\mcitedefaultmidpunct}
{\mcitedefaultendpunct}{\mcitedefaultseppunct}\relax
\EndOfBibitem
\bibitem[Bhakat and S{\"o}derhjelm(2017)Bhakat, and
  S{\"o}derhjelm]{Bhakatsampl5}
Bhakat,~S., and S{\"o}derhjelm,~P. (2017) Resolving the problem of trapped
  water in binding cavities: prediction of host--guest binding free energies in
  the SAMPL5 challenge by funnel metadynamics. \emph{Journal of Computer-Aided
  Molecular Design} \emph{31}, 119--132\relax
\mciteBstWouldAddEndPuncttrue
\mciteSetBstMidEndSepPunct{\mcitedefaultmidpunct}
{\mcitedefaultendpunct}{\mcitedefaultseppunct}\relax
\EndOfBibitem
\bibitem[Pant \latin{et~al.}(2020)Pant, Smith, Wang, Tajkhorshid, and
  Tiwary]{ravesgoop}
Pant,~S., Smith,~Z., Wang,~Y., Tajkhorshid,~E., and Tiwary,~P. (2020)
  Confronting pitfalls of AI-augmented molecular dynamics using statistical
  physics. \emph{The Journal of Chemical Physics} \emph{153}, 234118\relax
\mciteBstWouldAddEndPuncttrue
\mciteSetBstMidEndSepPunct{\mcitedefaultmidpunct}
{\mcitedefaultendpunct}{\mcitedefaultseppunct}\relax
\EndOfBibitem
\bibitem[Pfaendtner and Bonomi(2015)Pfaendtner, and Bonomi]{pbmetad}
Pfaendtner,~J., and Bonomi,~M. (2015) Efficient Sampling of High-Dimensional
  Free-Energy Landscapes with Parallel Bias Metadynamics. \emph{Journal of
  Chemical Theory and Computation} \emph{11}, 5062--5067\relax
\mciteBstWouldAddEndPuncttrue
\mciteSetBstMidEndSepPunct{\mcitedefaultmidpunct}
{\mcitedefaultendpunct}{\mcitedefaultseppunct}\relax
\EndOfBibitem
\bibitem[Weininger \latin{et~al.}(2014)Weininger, Modig, and Akke]{akkeflip}
Weininger,~U., Modig,~K., and Akke,~M. (2014) Ring Flips Revisited: 13C
  Relaxation Dispersion Measurements of Aromatic Side Chain Dynamics and
  Activation Barriers in Basic Pancreatic Trypsin Inhibitor.
  \emph{Biochemistry} \emph{53}, 4519--4525\relax
\mciteBstWouldAddEndPuncttrue
\mciteSetBstMidEndSepPunct{\mcitedefaultmidpunct}
{\mcitedefaultendpunct}{\mcitedefaultseppunct}\relax
\EndOfBibitem
\bibitem[Kulkarni and S{\"o}derhjelm(2021)Kulkarni, and
  S{\"o}derhjelm]{mandarflip}
Kulkarni,~M., and S{\"o}derhjelm,~P. (2021) Free Energy Landscape and Rate
  Estimation of the Aromatic Ring Flips in Basic Pancreatic Trypsin Inhibitor
  Using Metadynamics. \emph{bioRxiv} \relax
\mciteBstWouldAddEndPunctfalse
\mciteSetBstMidEndSepPunct{\mcitedefaultmidpunct}
{}{\mcitedefaultseppunct}\relax
\EndOfBibitem
\bibitem[Yu \latin{et~al.}(2020)Yu, Ma, Ren, Zhao, and Yi]{transformer}
Yu,~C., Ma,~X., Ren,~J., Zhao,~H., and Yi,~S. Spatio-Temporal Graph Transformer
  Networks for Pedestrian Trajectory Prediction. 2020\relax
\mciteBstWouldAddEndPuncttrue
\mciteSetBstMidEndSepPunct{\mcitedefaultmidpunct}
{\mcitedefaultendpunct}{\mcitedefaultseppunct}\relax
\EndOfBibitem
\end{mcitethebibliography}

%% graphical table of contents.
%%%%%%%%%%%%%%%%%%%%%%%%%%%%%%%%%%%%%%%%%%%%%%%%%%%%%%%%%%%%%%%%%%%%%

\end{document}

\maketitle

Table 1. Torsional features and corresponding feature indexes used in TICA (Figure 7) and train classifier algorithms (Figure 14)
\begin{tcolorbox}[breakable, size=fbox, boxrule=.5pt, pad at break*=1mm, opacityfill=0]
\label{tablesvm}
\begin{Verbatim}[commandchars=\\\{\}]
featureindex  atominds featuregroup featurizer otherinfo resids resnames  
0       [15, 17, 19, 22]         chi1   Dihedral       sin   [75]    [GLU]
1       [30, 32, 34, 37]         chi1   Dihedral       sin   [76]    [MET]
2       [47, 49, 51, 54]         chi1   Dihedral       sin   [77]    [ASN]
3       [61, 63, 65, 68]         chi1   Dihedral       sin   [78]    [TYR]
4       [82, 84, 86, 88]         chi1   Dihedral       sin   [79]    [VAL]
5    [98, 100, 102, 105]         chi1   Dihedral       sin   [80]    [SER]
6   [116, 118, 120, 126]         chi1   Dihedral       sin   [82]    [THR]
7   [130, 132, 134, 136]         chi1   Dihedral       sin   [83]    [VAL]
8   [146, 148, 150, 153]         chi1   Dihedral       sin   [84]    [SER]
9       [15, 17, 19, 22]         chi1   Dihedral       cos   [75]    [GLU]
10      [30, 32, 34, 37]         chi1   Dihedral       cos   [76]    [MET]
11      [47, 49, 51, 54]         chi1   Dihedral       cos   [77]    [ASN]
12      [61, 63, 65, 68]         chi1   Dihedral       cos   [78]    [TYR]
13      [82, 84, 86, 88]         chi1   Dihedral       cos   [79]    [VAL]
14   [98, 100, 102, 105]         chi1   Dihedral       cos   [80]    [SER]
15  [116, 118, 120, 126]         chi1   Dihedral       cos   [82]    [THR]
16  [130, 132, 134, 136]         chi1   Dihedral       cos   [83]    [VAL]
17  [146, 148, 150, 153]         chi1   Dihedral       cos   [84]    [SER]
18      [17, 19, 22, 25]         chi2   Dihedral       sin   [75]    [GLU]
19      [32, 34, 37, 40]         chi2   Dihedral       sin   [76]    [MET]
20      [49, 51, 54, 55]         chi2   Dihedral       sin   [77]    [ASN]
21      [63, 65, 68, 69]         chi2   Dihedral       sin   [78]    [TYR]
22      [17, 19, 22, 25]         chi2   Dihedral       cos   [75]    [GLU]
23      [32, 34, 37, 40]         chi2   Dihedral       cos   [76]    [MET]
24      [49, 51, 54, 55]         chi2   Dihedral       cos   [77]    [ASN]
25      [63, 65, 68, 69]         chi2   Dihedral       cos   [78]    [TYR]
\end{Verbatim}
\end{tcolorbox}

    % Add a bibliography block to the postdoc